\documentclass{gNST2e_arxiv_version}
\usepackage{times}
\usepackage{amsmath, amsthm, amssymb, mathtools}
\usepackage{nicefrac}       
\usepackage{microtype}		
\usepackage[group-separator={,},group-minimum-digits={3}]{siunitx}
\newcommand{\relmiddle}[1]{\mathrel{}\middle#1\mathrel{}}
\theoremstyle{plain}
\newtheorem{theorem}{Theorem}[section]
\theoremstyle{remark}
\newtheorem{remark}{Remark}
\theoremstyle{definition}
\newtheorem{definition}{Definition}
\theoremstyle{remark}
\newtheorem{condition}{Condition}
\newtheorem{step}{Step}

\usepackage{graphicx}
\usepackage{subcaption}

\usepackage[singlelinecheck=false,justification=justified,labelsep=space]{caption}
\usepackage{multirow}

\usepackage[plain,noend]{algorithm2e}

\usepackage[dvipsnames]{xcolor}
\usepackage{hyperref}       
\hypersetup{
colorlinks=true,
linkcolor=Blue,
citecolor=Blue,
filecolor=Blue,      
urlcolor=Blue,
anchorcolor=Blue
}

\usepackage{natbib}

\usepackage[capitalize, nameinlink, noabbrev]{cleveref} 

\crefname{condition}{Condition}{Conditions}
\newcommand{\pkg}[1]{{\fontseries{m}\fontseries{b}\selectfont #1}}

\begin{document}

\title{Empirical likelihood for the analysis of experimental designs}

\author{
\name{Eunseop Kim\textsuperscript{a}\(^{\ast}\)\thanks{\(^\ast\)Corresponding author. Email: kim.7302@osu.edu},
Steven N. MacEachern\textsuperscript{a}
and Mario Peruggia\textsuperscript{a}}
\affil{\textsuperscript{a}Department of Statistics, The Ohio State University,
\\1958 Neil Ave., Columbus, Ohio 43210, U.S.A.}
}

\maketitle

\begin{abstract}
Empirical likelihood enables a nonparametric, likelihood-driven style of inference without restrictive assumptions routinely made in parametric models. 
We develop a framework for applying empirical likelihood to the analysis of experimental designs, addressing issues that arise from blocking and multiple hypothesis testing.
In addition to popular designs such as balanced incomplete block designs, our approach allows for highly unbalanced, incomplete block designs.
We derive an asymptotic multivariate chi-square distribution for a set of empirical likelihood test statistics and propose two single-step multiple testing procedures: asymptotic Monte Carlo and nonparametric bootstrap. 
Both procedures asymptotically control the generalized family-wise error rate and efficiently construct simultaneous confidence intervals for comparisons of interest without explicitly considering the underlying covariance structure.
A simulation study demonstrates that the performance of the procedures is robust to violations of standard assumptions of linear mixed models.
We also present an application to experiments on a pesticide.
\end{abstract}

\begin{keywords}
Block design; Bootstrap; Family-wise error rate; Multiple testing; Multivariate chi-square distribution.
\end{keywords}

\begin{classcode}
62G09; 62G10; 62G15; 62G20; 62G30.
\end{classcode}

\section{Introduction}
In designed experiments, questions of particular interest frequently involve differences in means of a set of treatments and multiple comparisons. 
Classical parametric tools for analysis, such as the \({F}\)-test, Tukey test, or Ryan/Einot--Gabriel/Welsch test, provide efficient ways for testing hypotheses and constructing simultaneous confidence intervals (SCIs) but rely on restrictive assumptions on underlying distributions, variances, and sample sizes. 
Issues of misspecification and robustness of inference arise when these assumptions are not met.
In randomized block designs, rank-based, distribution-free multiple comparison procedures have been suggested, going back to \citet{friedman1937use} and \citet{nemenyi1963distribution}.
\cite{mansouri2004nonparametric} developed a Tukey-type nonparametric pairwise comparison procedure for balanced incomplete block designs.
More recently, \citet{eisinga2017exact} proposed an exact test for simultaneous pairwise comparison of Friedman rank sums with a method to quickly calculate the exact \({p}\)-values and associated statistics.
Rank-based approaches, however, have limitations in that they do not fully utilize the available data and have a well-known cycling inconsistency issue \citep{lehmann1975nonparametrics,fey2012consistency}.

Empirical likelihood \citep{owen1988empirical} can be helpful as a nonparametric alternative in such situations.
With suitably defined estimating functions, empirical likelihood enables nonparametric, likelihood-driven inferences without distributional specifications. 
It is well established that various forms of empirical likelihood ratio functions admit a nonparametric version of Wilks' theorem under mild conditions, providing a basis for an asymptotic test based on a chi-square null distribution; see, e.g.~\citet{qin1994empirical} and \citet{owen2001empirical}.
In addition, the empirical distribution of the data determines the shape and orientation of confidence regions. 
The coverage accuracy of the confidence regions can further be improved by bootstrap or Bartlett-correction \citep{diciccio1991empirical}.
In the context of the analysis of designed experiments, empirical likelihood has been studied for inference on the median using ranking data by \citet{liu2012rank} and \citet{alvo2015empirical}.
Inference on the mean is also available by formulating an appropriate estimating function.
Designs without a blocking factor, for example, can be analyzed as an analysis of variance problem \citep{owen1991empirical}. 
Popular block designs such as randomized complete block designs or balanced incomplete block designs can also be reconfigured as a multivariate mean problem.

The existing literature has mainly focused on establishing limit theorems for a single empirical likelihood (ratio) statistic with a single hypothesis.
\citet{wang2018f} applied \({F}\)-distribution calibrated empirical likelihood statistics to multiple hypothesis tests, assuming independence between tests. 
However, these results cannot be directly extended to various dependence scenarios, including the problem of multiple comparisons.
Although individual \({p}\)-values from empirical likelihood tests can be substituted into many existing multiple testing procedures, constructing SCIs for the comparisons based on empirical likelihood has not yet been investigated.
This article addresses the challenges of the multiplicity of comparisons by introducing an asymptotic framework for general block designs that leads to manageable inference. 
In particular, each confidence interval has a variable length that accommodates the underlying covariance structure without explicit studentization, and the SCIs achieve the target coverage probability asymptotically.
We also propose empirical likelihood-based multiple testing procedures that rest on this framework.
These procedures are generally applicable to other models and estimating functions.

The article is organized as follows. 
\cref{sec:preliminaries} introduces some preliminary concepts and conditions used in the rest of the article.
\cref{sec:theory} develops an asymptotic theory for a set of empirical likelihood test statistics. 
We propose two multiple testing procedures in \cref{sec:procedure} and evaluate the performances of the procedures in \cref{sec:simulation} through a simulation study.
An application to pesticide concentration experiments is discussed in Section \cref{sec:application}. 
We conclude with a discussion of directions for future research in \cref{sec:discussion}. 
The proofs of the theoretical results are provided in \hyperref[sec:appendix]{Appendix}.

\section{Preliminaries}\label{sec:preliminaries}
\subsection{General block designs}\label{subsec:gbd}
A block design is an ordered pair \({(\mathcal{T}, \mathcal{B})}\) where \({\mathcal{T}}\) is a set of \({p}\) points that we call treatments, and \({\mathcal{B}}\) is a collection of \({n}\) nonempty subsets of \({\mathcal{T}}\) called blocks. 
We consider general block designs where each block size is \({b_i}\), with \({1} \leq {b_i} \leq {p}\), for \({i} = {1, \dots, n}\). 
Treatment \({k}\) is contained in \({r_k}\) blocks and each pair of distinct treatments \({k}\) and \({l}\) is contained in \({\lambda_{kl}}\) blocks, for \({k, l} = {1, \dots, p}\). 
Then we have the following set of equations:
\begin{equation*}
\sum_{i = 1}^n b_i = \sum_{k = 1}^p r_k
\textnormal{ and } 
\sum_{l \neq k}\lambda_{kl} = \sum_{i \in \mathcal{B}_k}(b_i - 1) \textnormal{ for each } k,
\end{equation*}
where \({\mathcal{B}_k} \subseteq {\{1, \dots, n\}}\) denotes the index set of the blocks containing treatment \({k}\). 
Let \({C_n}\) denote the associated \({n \times p}\) binary incidence matrix with the \({(i, k)}\) component given by \({c_{ik} = 1(i \in \mathcal{B}_k)}\), where \({1(\cdot)}\) is the indicator function of its argument. 
The \({i}\)th row sum is then \({b_i}\) and the \({k}\)th column sum is \({r_k}\). 

The \({n}\) blocks are regarded as random samples from an unknown population.
Specifically, we assume independent and identically distributed \({(\textnormal{i.i.d.})}\) \({p}\)-dimensional random variables \({X_1, \dots, X_n}\) defined on a probability space \({(\Omega, \mathcal{F}, P)}\) with mean \({E(X_1)} = {\theta_0} \in {\textnormal{int}(\theta)}\) and positive definite covariance matrix \({\textnormal{Var}(X_1)} = {\Sigma}\), where \({\Theta} \subseteq {\mathbb{R}^{p}}\) denotes the parameter space. 
The parameter of interest is \({\theta_0}\), the treatment effects. 
According to the design and \({C_n}\), we only observe those \({X_{ik}}\) from \({X_i} = {(X_{i1}, \dots, X_{ip})}\) for which \({c_{ik}} = {1}\). 
Since \({C_n}\) is always available, we do not make a notational distinction between the underlying random variable \({X_i}\) and its observable components. 
It will be clear from the context what we are referring to. 
In order to work with empirical likelihood, we require that \({n^{-1} C_n^\top C_n} \to {D}\) as \({n} \to {\infty}\) for some matrix \({D}\) with positive diagonal entries.

\subsection{Empirical likelihood for block designs}
We introduce the general setup for empirical likelihood within the block design framework. 
The available data are denoted by \({\mathcal{X}_n} = {\{X_1, \dots, X_n\}}\). 
Inference for \({\theta_0}\) is based on a \({p}\)-dimensional estimating function \({g(X_i, \theta)}\), where \({g(X_i, \theta)}\) equals \({X_i - \theta}\) with the unobserved components set to \({0}\). 
More explicitly, we write 
\begin{equation}\label{eq:estimating function}
g(X_i, \theta) 
\equiv 
g(X_i, \theta; c_i) 
= 
(X_i - \theta) \circ c_i,
\end{equation}
where \({c_i}\) is the \({i}\)th row of \({C_n}\) and \({\circ}\) is the Hadamard product. 
The (profile) empirical likelihood ratio function, evaluated at \({\theta}\), is defined as
\begin{equation*}
\max_{w_i}\left\{\prod_{i = 1}^n nw_i : w_i > 0, \sum_{i = 1}^n w_i = 1, \textnormal{ and } \sum_{i = 1}^n w_i g(X_i, \theta) = 0 \right\}.
\end{equation*}
A unique solution exists if the zero vector is contained in \({\textnormal{Conv}_n(\theta)}\), where \({\textnormal{Conv}_n(\theta)}\) denotes the interior of the convex hull of \({\{g(X_i, \theta): i = 1, \dots, n\}}\).
The Lagrange multipliers \({\lambda} \equiv {\lambda(\theta)}\) of the dual optimization problem solve
\begin{equation*}
\frac{1}{n}\sum_{i = 1}^n \frac{g(X_i, \theta)}{1 + \lambda^\top g(X_i, \theta)} = 0.
\end{equation*}
We denote minus twice the log empirical likelihood ratio function by
\begin{equation*}
l_n(\theta) = 2\sum_{i = 1}^n \log \left(1 + \lambda^\top g(X_i, \theta)\right).
\end{equation*}

In the case of \({g(X_i, \theta)} = {X_i - \theta}\), \citet{owen1990empirical} showed that \({l_n(\theta_0)}\) converges in distribution to \({\chi^2_p}\), a chi-square distribution with \({p}\) degrees of freedom.
Similar results also hold for other forms of estimating functions \citep{qin1994empirical}, and it can be shown that \({l_n(\theta_0)} \to {\chi^2_p}\) in distribution for our general block designs under some regularity conditions. 
A confidence region for \({\theta_0}\) can then be constructed as \({\{\theta: {l_n(\theta) \leq \chi^2_{p, \alpha}}\}}\), where \({\chi^2_{p, \alpha}}\) is the \({(1 - \alpha)}\)th quantile of a \({\chi^2_p}\) distribution. 

For the case of a subset of the parameter vector, let \({\theta} = {(\theta_{1}, \theta_{2})}\) for a \({q}\)-dimensional parameter \({\theta_{1}}\) with \({q} \leq {p}\), and consider testing a hypothesis \({H: {\theta_{1} = \theta_{1}^*}}\).
Under additional assumptions, a relevant test statistic \({l_n(\theta_{1}^*, \widehat{\theta}_{2})} \to {\chi^2_q}\) in distribution, where \({\widehat{\theta}_{2}}\) minimizes \({l_n(\theta_{1}^*, \theta_{2})}\) with respect to \({\theta_{2}}\) \citep[Corollary 5]{qin1994empirical}. 
More generally, for a \({q}\)-dimensional constraint \({h(\theta)} = {0}\), \citet{qin1995estimating} showed that if \({h(\theta_0)} = {0}\) then \({l_n(\widehat{\theta})} \to {\chi^2_q}\) in distribution, where \({\widehat{\theta}}\) denotes the minimizer of the problem.
\citet{adimari2010note} extended hypothesis testing with empirical likelihood to show that the chi-square calibration holds for an even broader class of estimating functions.
We can apply these results to general block designs and perform some important tests, including the test of no treatment effect or the interaction between treatments and the blocking variable. 
The applicability, however, is still restricted to a \textit{single} hypothesis test.

\subsection{Multiple testing}
Consider simultaneously testing \({m}\) null hypotheses \({H_j}\), \({j} = {1, \dots, m}\).
We assume that each \({H_j}\) corresponds to a nonempty subset of \({\theta}\) through a smooth \({q_j}\)-dimensional function \({h_j}\) such that \({H_j} = {\{\theta \in \Theta: h_j(\theta) = 0\}}\). 
We have \({\theta_0} \in {H_j}\) under \({H_j}\) (when \({H_j}\) is true). 
The complete null hypothesis \({H_0} = {\cap_j H_j}\) is also assumed to be nonempty. 
Then we denote a multiple testing procedure by \({\phi} = {\{\phi_j: j = 1, \dots, m\}}\), where \({\phi_j}\) maps the data \({\mathcal{X}_n}\) into \({\{0, 1\}}\), and \({H_j}\) is rejected if and only if \({\phi_j} = {1}\). 
We restrict our attention to procedures that provide a common cutoff value \({c_\alpha}\) at a nominal level \({\alpha} \in {(0, 1)}\). 
Given a vector of \({m}\) test statistics \({T_n} = {(T_{n1}, \dots, T_{nm})}\), we reject \({H_j}\) if \({T_{nj}} > {c_\alpha}\).
The total number of false rejections is \({V_m} = {\sum_{j \in \mathcal{I}_0} 1(\phi_j = 1)}\), where \({\mathcal{I}_0} = {\{j: \theta_0 \in H_j\}}\) is the index set of true null hypotheses. 

Of various Type \MakeUppercase{\romannumeral 1} error rates for multiple testing, the most common choice in designed experiments is family-wise error rate (FWER).
When the number of hypotheses is large, one can consider the generalized family-wise error rate (gFWER) as a less stringent alternative, which is defined as the probability of \({v}\) or more false rejections for some \({v} \leq {m}\). 
A discussion of procedures for gFWER control can be found in \citet{lehmann2005generalizations}.
Formally, a procedure \({\phi}\) is said to control gFWER (strongly) at level \({\alpha}\) if
\begin{equation*}
\textnormal{gFWER}_{\theta}(\phi) = P_{\theta}(V_m \geq v) \leq \alpha \textnormal{ for all } \theta \in \Theta.
\end{equation*}
When \({v} = {1}\), this reduces to FWER control. 
We say that \({\phi}\) controls gFWER asymptotically if \({\limsup_{n \to \infty}
\textnormal{gFWER}_{\theta}(\phi)} \leq {\alpha} {\textnormal{ for all }} {\theta} \in {\Theta}\).
This article addresses single-step procedures for gFWER control with consideration of the \textit{joint} distribution of the empirical likelihood statistics.

\section{Asymptotics for multiple testing}\label{sec:theory}
\subsection{Multivariate chi-square calibration}\label{subsec:mvchisq}
In order to address the multiplicity of our problem and formalize asymptotic multiple testing procedures based on empirical likelihood statistics, we first need a multivariate generalization of chi-square calibration, a multivariate chi-square distribution. 
The class of multivariate distributions with marginal chi-square distributions is much too broad to be useful in practice, and there is no universal definition of a multivariate chi-square distribution.

In what follows, we adopt a particular type of multivariate chi-square distribution introduced in \citet{dickhaus2014simultaneous}. 
\begin{definition}[\citet{dickhaus2014simultaneous}]
For a vector of positive integers \({q} = {(q_1, \dots, q_m)}\), let \({Z_j} = {(Z_{j1}, \dots, Z_{jq_j})} \sim {N(0, I_{q_j})}\), \({j} = {1, \dots, m}\). 
Assume that \({(Z_1, \dots, Z_m)}\) has a multivariate normal distribution with \({\sum_{j = 1}^m q_j \times \sum_{j = 1}^m q_j}\) correlation matrix 
\begin{equation*}
R = 
\left\{
\rho\left(Z_{j_1, l_1}, Z_{j_2, l_2}\right): j_1, j_2 = 1, \dots, m; l_1 = 1, \dots, q_{j_1}; l_2 = 1, \dots, q_{j_2}
\right\}. 
\end{equation*}
Let \({T} = {(T_1, \dots, T_m)}\), with \({T_j} = {Z_j^\top Z_j} \sim {\chi^2_{q_j}}\).
Then \({T}\) has a multivariate (central) chi-square distribution (of generalized Wishart-type) with parameters \({m}\), \({q}\), and \({R}\).
We write \({T} \sim {\chi^2(m, q, R)}\). 
\end{definition}
This distribution naturally arises as a joint limiting distribution of many Wald-type statistics and allows for varying degrees of freedom in each marginal. 
A comprehensive overview of different types of multivariate chi-square distributions and their applications can be found in \citet{dickhaus2015survey}. 

We now establish a multivariate extension that covers general block designs as a special case. 
To this end, we do not require \({\textnormal{i.i.d.}}\) observations \({X_i}\) and we allow the \({p}\)-dimensional estimating function \({g(X_i, \theta)}\) to take forms different from \eqref{eq:estimating function}.
Let \({\theta}\) be a parameter of interest (not necessarily the mean parameter) and define
\begin{equation*}
G(\theta) = E\left\{g(X_i, \theta)\right\},\
G_n(\theta) = \frac{1}{n}\sum_{i = 1}^n g(X_i, \theta),
\textnormal{ and } 
S_n(\theta) = \frac{1}{n}\sum_{i = 1}^n g(X_i, \theta) g(X_i, \theta)^\top,
\end{equation*}
with the property that \({G(\theta_0)} = {0}\).
Here and throughout, we use \({\vert\cdot\vert}\) to denote the Euclidean norm for vectors.
For matrices, \({\Vert\cdot\Vert}\) and \({\partial_{\theta}(\cdot)}\) denote the Frobenius norm and the Jacobian matrix, respectively.
All limits are taken as \({n} \to {\infty}\). 
We assume the following regularity conditions:
\begin{condition}\label{condition:convex hull}
\({P\left\{0 \in \textnormal{Conv}_n\left(\theta_0\right)\right\}} \to {1}\).
\end{condition}

\begin{condition}\label{condition:uniform consistency}
\({g(X_i, \theta)}\) and \({G(\theta)}\) are continuously differentiable in a neighborhood \({\mathcal{N}}\) of \({\theta_0}\) almost surely, and
\({\sup_{\theta \in \mathcal{N}} \Vert\partial_{\theta} G_n(\theta) - \partial_{\theta}G(\theta)\Vert} \to {0}\) in probability with nonsingular \({\partial_{\theta}G(\theta_0)}\).
\end{condition}

\begin{condition}\label{condition:consistent covariance estimation}
There exists a matrix function \({V(\theta)}\) with positive definite \({V(\theta_0)}\) such that \({\sup_{\theta \in \mathcal{N}} \Vert S_n(\theta) - V(\theta)\Vert} \to {0}\) in probability and \({\sup_{\vert\theta - \theta_0\vert \leq b_n} \Vert V(\theta) - V(\theta_0)\Vert} \to {0}\), for any sequence of positive real numbers \({b_n} \to {0}\). 
\end{condition}

\begin{condition}\label{condition:asymptotic normality}
\({a_n G_n(\theta_0)} \to {U}\) in distribution for a sequence of positive real numbers \({a_n} \to {\infty}\), where \({U} \sim {N(0, V(\theta_0))}\).
\end{condition}

\begin{condition}\label{condition:maximum bound}
\({\max_{1 \leq i \leq n} \vert g(X_i, \theta_0)\vert} = {o_{P}(a_n)}\) and 
\({\max_{1 \leq i \leq n} \Vert\partial_{\theta}{g}(X_i, \theta_0)\Vert} = {O_{P}(a_n)}\).
\end{condition}

\begin{condition}\label{condition:hypothesis}
The function \({H_j}\) defining the null hypothesis \({H_j}\) is continuously differentiable on \({\mathcal{N}}\) with Jacobian matrix \({J_j} = {\partial_{\theta} h_j(\theta_0)}\) of full rank \({q_j} \leq {p}\), \({j} = {1, \dots, m}\).
\end{condition}

\cref{condition:convex hull} is the basic existence condition for empirical likelihood in the asymptotic setting. 
Since the computation of \({l_n(\theta)}\) involves the quadratic forms \({G_n}\) and \({S_n}\), \cref{condition:uniform consistency,condition:consistent covariance estimation} are required for \({l_n(\theta)}\), as a smooth function of \({\theta}\), to be evaluated in a neighborhood of \({\theta_0}\). 
\cref{condition:asymptotic normality} implies that, asymptotically, the quadratic forms have marginal and joint multivariate chi-square distributions.
\cref{condition:maximum bound} demands that the remainder terms be negligible.
\cref{condition:hypothesis} completes the statement of the constrained empirical likelihood problems for multiple testing and holds for most practical applications that we consider.

For \({j} = {1, \dots, m}\), we define the empirical likelihood statistic associated with hypothesis \({H_j}\) as 
\begin{equation}\label{eq:EL statistic}
T_{nj} = \frac{a_n^2}{n} \inf_{\theta \in H_j \cap \overline{K}_n} l_n(\theta),
\end{equation}
where \({\overline{K}_n} = {\{\theta: \vert\theta - \theta_0\vert \leq K/a_n\}}\) denotes a sequence of closed balls around \({\theta_0}\) for a given \({K} > {0}\).
The \({m}\)-dimensional test statistic is denoted by \({T_n} = {(T_{n1}, \dots, T_{nm})}\). 
For notational convenience, let \({V} = {V(\theta_0)}\), \({W} =
{\partial_{\theta}G(\theta_0)^{-1}}\), and \({M} = {W V W^\top}\).
Then, define \({T} = {(T_1, \dots, T_m)}\), where \({T_j} = {U^\top A_j U}\) with \({A_j} = {(J_j W)^\top (J_j M {J_j}^\top)^{-1}(J_j W)}\).
\begin{theorem}\label{thm:mvchisq}
Under \({H_0}\) and \cref{condition:convex hull,condition:uniform consistency,condition:asymptotic normality,condition:maximum bound,condition:consistent covariance estimation,condition:hypothesis}, 
\begin{equation*}
T_n \to T \sim \chi^2(m, q, R)
\end{equation*}
in distribution for some sequence \({\overline{K}_n}\).
Here, \({q} = {(q_1, \dots, q_m)}\) and \({R}\) is the correlation matrix of \({(Z_1, \dots, Z_m)}\), with \({Z_j} = {(J_j M {J_j}^\top)^{-1/2} J_j W U}\).
\end{theorem}

\begin{remark}\label{rmk:gbd}
For the general block designs introduced in \cref{subsec:gbd}, it can be shown that \({l_n}\) is convex in \({\theta}\) so we can find a solution \({\widehat{\theta}_c}\) of the optimization problem in \eqref{eq:EL statistic} such that \({\widehat{\theta}_c - \theta_0} = {O_{P}(a_n^{-1})}\).
Thus, the closed ball constraint is not binding asymptotically, i.e.~\({a_n^2 n^{-1}\inf_{\theta \in H_j} l_n(\theta)} = {a_n^2 n^{-1}\inf_{\theta: H_j \cap \overline{K}_n} l_n(\theta)} + {o_{P}(1)}\). 
In other cases with general estimating functions, identification of \({\theta_0}\) may require additional conditions, such as compactness of \({\theta}\), or \({\theta_0}\) being the unique zero of \({G(\theta)}\) \citep[see, e.g.][]{yuan1998asymptotics,jacod2018review}.
\end{remark}

\begin{remark}\label{rmk:subsetpivot}
\({T_n}\) satisfies the so-called subset pivotality condition \citep{westfall1993resampling} asymptotically in the sense that, for any subset \({\mathcal{S}} \subseteq {\{1, \dots, m\}}\) the joint limiting distribution of \({\{T_{nj}: j \in \mathcal{S}\}}\) remains the same under \({\cap_{j \in \mathcal{S}} H_j}\) and \({H_0}\).
\end{remark}

\subsection{Illustration of the theory for general block designs}\label{subsec:illustration}
We give an illustration of the preceding theory by verifying \cref{condition:convex hull,condition:uniform consistency,condition:asymptotic normality,condition:maximum bound,condition:consistent covariance estimation} of \cref{thm:mvchisq} for general block designs. 
\cref{condition:convex hull} holds by applying the Glivenko--Cantelli argument over the half-spaces of \citet[p.~219]{owen2001empirical}.
\cref{condition:uniform consistency} is checked by noting that both \({g(X_i, \theta)}\) and \({G(\theta)}\) are continuously differentiable with \({\partial_{\theta}G_n(\theta)} = 
{-n^{-1}\textnormal{diag}(r_1, \dots, r_p)}\) and \({\partial_{\theta}G(\theta)} = {-\textnormal{diag}(D)}\),
where \({\textnormal{diag}(\cdot)}\) denotes the diagonal matrix of its argument (either a vector or a matrix). 
The result follows since we have assumed that \({n^{-1}C_n^\top C_n} \to {D}\). 
For \cref{condition:consistent covariance estimation}, observe that
\begin{align*}
\Vert S_n\left(\theta\right) -& V\left(\theta\right)\Vert 
\leq \\
&\Vert S_n\left(\theta\right) - S_n\left(\theta_0\right) - \left(\theta - \theta_0\right)\left(\theta - \theta_0\right)^\top \circ D\Vert 
+ 
\left\Vert S_n\left(\theta_0\right) - V_n\right\Vert +
\left\Vert V - V_n\right\Vert,
\end{align*}
where
\begin{align*}
S_n(\theta) -& S_n(\theta_0) - (\theta - \theta_0)(\theta - \theta_0)^\top \circ D
= 
\frac{1}{n}\sum_{i = 1}^n
(X_i - \theta_0)(\theta_0 - \theta)^\top
\circ
c_i c_i^\top\\
&+
\frac{1}{n}\sum_{i = 1}^n
(\theta_0 - \theta)(X_i - \theta_0)^\top
\circ
c_i c_i^\top + 
(\theta - \theta_0)(\theta - \theta_0)^\top
\circ
\left(\frac{1}{n}C_n^\top C_n - D\right)
\end{align*}
and
\begin{equation*}
\left\Vert S_n\left(\theta_0\right) - V_n\right\Vert^2 
= 
\sum_{k = 1}^p \sum_{l = 1}^p 
\left\vert
\frac{1}{n}
\sum_{i \in \mathcal{B}_k \cap \mathcal{B}_l}\left(X_{ik} - \theta_{0k}\right)\left(X_{il} - \theta_{0l}\right)
- \frac{\sigma_{kl}\lambda_{kl}}{n}
\right\vert^2,
\end{equation*}
with the \({(k, l)}\) component of \({\Sigma}\) denoted by \({\sigma_{kl}}\).
Then, we have 
\begin{equation*}
S_n(\theta) - S_n(\theta_0) - (\theta - \theta_0)(\theta - \theta_0)^\top \circ D \to 0
\textnormal{ and }
S_n(\theta_0) - V_n \to 0
\end{equation*}
almost surely uniformly in \({\theta} \in {\mathcal{N}}\) by the (uniform) law of large numbers, which verifies the first requirement of \cref{condition:consistent covariance estimation} such that \({\sup_{\theta \in \mathcal{N}} \vert S_n(\theta) - V(\theta)\vert} \to {0}\) in probability. 
For the second requirement,
\begin{equation*}
\sup_{\left\vert\theta - \theta_0\right\vert \leq b_n} \left\Vert V\left(\theta\right) - V\right\Vert
=
\sup_{\left\vert\theta - \theta_0\right\vert \leq b_n} \Vert\left(\theta - \theta_0\right)\left(\theta - \theta_0\right)^\top \circ D\Vert
\leq p^2 b_n^2 \to 0,
\end{equation*}
establishing \cref{condition:consistent covariance estimation}.
For \cref{condition:asymptotic normality}, we take \({a_n} = {\sqrt{n}}\) and choose any \({\epsilon} > {0}\). 
Let
\begin{equation*}
V_n = \frac{1}{n}\sum_{i = 1}^n \textnormal{Var}\left\{g(X_i, \theta_0)\right\} = E\left\{S_n(\theta_0)\right\} = \Sigma \circ \frac{1}{n}C_n^\top C_n.
\end{equation*}
We have \({V_n \to V = \Sigma \circ D}\) where \({V}\) is positive definite, and
\begin{align*}
\frac{1}{n}\sum_{i = 1}^n E&\left\{
\left\vert g(X_i, \theta_0)\right\vert^2 1\left(\left\vert g\left(X_i, \theta_0\right) \right\vert \geq \epsilon \sqrt{n}\right)\right\} \\
&\leq \frac{1}{n} \sum_{i = 1}^n E\left\{
\left\vert X_i - \theta_0\right\vert^2 1\left(\left\vert X_i - \theta_0 \right\vert \geq \epsilon \sqrt{n}\right)
\right\} \to 0.
\end{align*}
It follows from the Lindeberg--Feller central limit theorem that 
\({a_n G_n(\theta_0)} \to {N(0, V)}\) in distribution, and \cref{condition:asymptotic normality} holds with \({V(\theta)} = {V + (\theta - \theta_0)(\theta - \theta_0)^\top \circ D}\).
Next, a Borel--Cantelli argument \citep[Lemma 11.2]{owen2001empirical} shows that \({\max_{1\leq i \leq n} \vert X_i - \theta_0\vert} = {o_{P}(a_n)}\). 
Since \({\max_{1 \leq i \leq n} \Vert\partial_{\theta}{g}(X_i, \theta_0)\Vert} \leq {\sqrt{p}}\), we have
\({\max_{1 \leq i \leq n} \Vert\partial_{\theta}{g}(X_i, \theta_0)\Vert} = {o(a_n)}\) almost surely and \cref{condition:maximum bound} is checked.

\section{Empirical likelihood-based multiple testing}\label{sec:procedure}
\subsection{Asymptotic Monte Carlo}\label{subsec:AMC}
This section extends the multivariate empirical likelihood theory developed in \cref{sec:theory} to specific multiple testing procedures for general block designs. 
We propose two procedures for calibration of the common cutoff value, where the finite-sample null distribution of \({T_n}\) is approximated by employing appropriate schemes. 
Both procedures determine cutoffs that provide asymptotic gFWER control (see \cref{rmk:subsetpivot}).

As a multivariate analog of chi-square calibration, one may consider relying on multivariate chi-square quantiles of \({T}\) as a cutoff. 
In practice, however, the covariance matrix \({V}\) of \({U}\) and thus the correlation matrix \({R}\) of \({T}\) is rarely known, making it impossible to compute the multivariate quantiles directly. 
As an alternative, the asymptotic Monte Carlo (AMC) procedure relies on the stochastic representation in \cref{thm:mvchisq} to produce a simulation-based approximation to the distribution of \({T}\) up to any desired precision.

Suppose that we have a consistent estimator \({\widehat{\theta}}\) of \({\theta_0}\). 
It can be shown from \cref{condition:consistent covariance estimation} that \({S_n(\widehat{\theta})} \to {V}\) in probability; see \citet[Remark 2.2]{hjort2009extending}. 
Then, the AMC procedure consists of replacing \({V}\) with \({S_n(\widehat{\theta})}\) and simulating samples from the approximate distribution \({N(0, S_n(\widehat{\theta}))}\).
Let \({\widehat{A}_j} = 
{(J_j W)^\top
\{(J_j W) S_n(\widehat{\theta}) (J_j W)^\top\}^{-1}(J_j W)}\) and consider random variables \({U_n} \sim {N(0, S_n(\widehat{\theta}))}\) and \({\widehat{T}_n} = {(U_n^\top \widehat{A}_1 U_n, \dots, U_n^\top \widehat{A}_m U_n)}\), defined conditionally on the observed data \({\mathcal{X}_n}\). 
With \({P_n}\) denoting the conditional distribution of \({\widehat{T}_n}\), the following theorem ensures that the distance between \({P_n}\) and the distribution of \({T_n}\) converges to zero in probability. 

\begin{theorem}\label{thm:amc}
Under \({H_0}\) and \cref{condition:convex hull,condition:uniform consistency,condition:asymptotic normality,condition:maximum bound,condition:consistent covariance estimation,condition:hypothesis}, 
\begin{equation*}
\sup_{x \in \mathbb{R}^m_+} \left\vert 
P_n(\widehat{T}_n \leq x \mid \mathcal{X}_n)
-
P\left(T_n \leq x \right)
\right\vert
\to 0
\end{equation*}
in probability.
\end{theorem}
\cref{thm:amc} guarantees the asymptotic validity of the AMC procedure described in \cref{alg:AMC}. 
For \({v} = {1}\), the procedure reduces to controlling the asymptotic FWER and the cutoff \({c_\alpha^\textnormal{AMC}}\) is computed from the maximum statistics \({\{\widehat{Q}_{n(m)}^{(1)},\dots, \widehat{Q}_{n(m)}^{(B)}\}}\). 
When the \({H_j}\)s are contrasts of the form \({\sum_{k = 1}^p u_k \theta_k}\) with known constants \({u_1, \dots, u_p}\), asymptotic \({100(1 - \alpha)\%}\) SCIs for the \({H_j}\)s can be \({\{r \in \mathbb{R}:\inf_{h_j(\theta) = r}l_n(\theta) \leq {c_\alpha^\textnormal{AMC}}\}}\). 
The test procedure and the SCIs are compatible, i.e.~whenever \({H_j}\) is rejected, the corresponding interval does not include the null value and vice versa.
\begin{algorithm}[t]
\caption{AMC}\label{alg:AMC}
\DontPrintSemicolon
\SetNlSty{textbf}{}{.}
\KwData{\({\mathcal{X}_n}\)}
\KwResult{cutoff \({c_\alpha}\) and adjusted \({p}\)-values \({\widetilde{p}_1, \dots, \widetilde{p}_m}\)}
\BlankLine
\nl Compute \({T_n}\), \({S_n(\widehat{\theta})}\), and \({\widehat{A}_1, \dots, \widehat{A}_m}\) \; 
\BlankLine
\nl Monte Carlo simulation for approximation\;
\For{\({b} = {1, \dots, B}\)}{
Simulate \({U_n^{(b)} \sim N(0, S_n(\widehat{\theta}))}\)\;
\BlankLine
\({\widehat{Q}^{(b)}_{n(v)} \leftarrow }\) the \({v}\)th largest component of \({\left({U_n^{(b)}}^\top \widehat{A}_1 U_n^{(b)}, \dots, {U_n^{(b)}}^\top \widehat{A}_m U_n^{(b)}\right)}\) \;
}
\BlankLine
\nl \({c_\alpha} \leftarrow {(1 - \alpha)}\)th quantile of \({\left\{\widehat{Q}_{n(v)}^{(1)} ,\dots, \widehat{Q}_{n(v)}^{(B)}\right\}}\) \;
\BlankLine
\nl Adjusted \({p}\)-values and multiple testing\;
\For{\({j} = {1, \dots, m}\)}{
\begin{flalign*}
&\widetilde{p}_j \leftarrow \frac{1}{B}\sum_{b = 1}^B 1\left(\widehat{Q}_{n(v)}^{(b)} \geq T_{nj}\right)&
\end{flalign*}
\BlankLine
\lIf{
\({T_{nj}} > {c_\alpha} \textnormal{ or } {\widetilde{p}_j} < {\alpha}\)} 
{reject \({H_j}\)}
}
\end{algorithm}

\begin{remark}
Rather than generating draws from an approximate multivariate chi-square distribution, low-dimensional multiplicity-adjusted
quantiles can be computed numerically if the underlying correlation matrix fulfills certain structural properties \citep{stange2016computing}. 
\end{remark}

\subsection{Nonparametric bootstrap}\label{subsec:NB}
It has been widely noted that the error rates of tests based on the asymptotic chi-square calibration tend to be higher than the nominal levels, especially in small sample or high-dimensional problems; see, e.g.~\citet{qin1994empirical} and \citet{tsao2004bounds}. 
This issue persists in our setting with multiple empirical likelihood statistics.
Moreover, considering the incomplete nature of block designs, convergence to a multivariate chi-square distribution may be slow. 
As an alternative, \citet{owen1988empirical} proposed a bootstrap calibration for the mean.
Resampling from the original data \({\mathcal{X}_n}\) yields bootstrap replicates \({\mathcal{X}_n^{(b)}}\), \({b} = {1, \dots, B}\). 
For each \({\mathcal{X}_n^{(b)}}\), the empirical likelihood statistic \({l_n^{(b)}(\overline{X})}\) is computed at the sample mean \({\overline{X}}\) of \({\mathcal{X}_n}\). 
The cutoff is obtained as the sample \({(1 - \alpha)}\)th quantile of \({\{l_n^{(1)}(\overline{X}), \dots, l_n^{(B)}(\overline{X})\}}\). 

In our setting, let \({\widetilde{\mathcal{X}}_n}\) be the null-transformed data with \({H_0}\) imposed on the observed data \({\mathcal{X}_n}\) (see \eqref{eq:null transformation} below as an example).
Then we denote the (nonparametric) bootstrap samples by \({\widetilde{\mathcal{X}}_n^*} = {\{\widetilde{X}_1^*, \dots, \widetilde{X}^*_n\}}\), where \({\widetilde{X}_i^*}\), \({i} = {1, \dots, n}\), are \({\textnormal{i.i.d.}}\) observations from \({\widetilde{\mathcal{X}}_n}\). 
Conditional on \({\widetilde{\mathcal{X}}_n^*}\), we denote the bootstrap empirical likelihood statistic by \({l^*_n(\theta)}\) and the test statistics by \({T^*_n} = {(T^*_{n1}, \dots, T^*_{nm})}\), where \({T^*_{nj}} = {a_n^2n^{-1}\inf_{\theta \in H_j} l_n^*(\theta)}\). 
We establish another consistency result that provides the weak convergence of \({T_n^*}\) in probability to \({T}\). 
As is customary, we denote the bootstrap distribution conditional on the data by \({P_n^*}\).

\begin{theorem}\label{thm:nb}
Under \({H_0}\) and \cref{condition:convex hull,condition:uniform consistency,condition:asymptotic normality,condition:maximum bound,condition:consistent covariance estimation,condition:hypothesis}, if \({E(\vert X_i \vert^4)} < {\infty}\),
\begin{equation*}
\sup_{x \in \mathbb{R}^m_+} \left\vert 
P_n^*\left(T_n^* \leq x \mid \mathcal{X}_n
\right)
-
P\left(T_n \leq x \right)
\right\vert
\to 0
\end{equation*}
in probability.
\end{theorem}
\cref{thm:nb} ensures that the conditional distribution of \({T^*_n}\) approximates the multivariate chi-square distribution of \({T}\). 
Adding the continuity of \({T}\) implies that the procedures for gFWER control can be asymptotically calibrated by the bootstrap replicates of \({T_n^*}\), namely \({T_n^{(1)}, \dots, T_n^{(B)}}\). 

In \cref{alg:NB} we describe the nonparametric bootstrap (NB) procedure. 
It differs from the AMC procedure only in the cutoff \({c_\alpha^{\textnormal{NB}}}\) and the resulting adjusted \({p}\)-values.
Our experience with the procedure shows that the NB procedure is better tuned to the distribution from which the data arise and that \({c_\alpha^{\textnormal{NB}}}\) is typically larger than \({c_\alpha^{\textnormal{AMC}}}\).
\begin{algorithm}[t]
\caption{NB}\label{alg:NB}
\DontPrintSemicolon
\SetNlSty{textbf}{}{.}
\KwData{\({\mathcal{X}_n}\) and \({\widetilde{\mathcal{X}}_n}\)}
\KwResult{cutoff \({c_\alpha}\) and adjusted \({p}\)-values \({\widetilde{p}_1, \dots, \widetilde{p}_m}\)}
\BlankLine
\nl Compute \({T_n}\) \; 
\BlankLine
\nl Bootstrapping for approximation\;
\For{\({b} = {1, \dots, B}\)}{
Simulate \({\widetilde{\mathcal{X}}_n^{(b)}}\) from \({\widetilde{\mathcal{X}}_n}\) and compute \({T_n^{(b)}}\) \;
\BlankLine
\({T_{n(v)}^{(b)} \leftarrow}\) the \({v}\)th largest component of \({T_n^{(b)}}\) \;
}
\BlankLine
\nl \({c_\alpha} \leftarrow {(1 - \alpha)}\)th quantile of \({\left\{T_{n(v)}^{(1)} ,\dots, T_{n(v)}^{(B)}\right\}}\) \;
\BlankLine
\nl Adjusted \({p}\)-values and multiple testing\;
\For{\({j} = {1, \dots, m}\)}{
\begin{flalign*}
&\widetilde{p}_j \leftarrow \frac{1}{B}\sum_{b = 1}^B 1\left(T_{n(v)}^{(b)} \geq T_{nj}\right)&
\end{flalign*}
\BlankLine
\lIf{\({T_{nj}} > {c_\alpha} \textnormal{ or } {\widetilde{p}_j < \alpha}\)} 
{reject \({H_j}\)}
}
\end{algorithm}

\begin{remark}\label{rmk:nb}
Other bootstrap schemes can be considered as well. 
The block bootstrap, for instance, may be adapted to produce bootstrap replicates that better preserve the original design structure. 
This can be of great importance when \({n}\) is small and the convex hull constraint is of concern. 
In this regard, it is worth examining the applicability of alternative formulations of empirical likelihood that are free from the constraint \citep[see, e.g.][]{chen2008adjusted, tsao2013empirical}.
\end{remark}

\section{Simulation study}\label{sec:simulation}
In this section, we carry out a simulation study on a balanced incomplete block design for all pairwise comparisons of treatment effects.
The design has five treatments with \({n}\) blocks, and each block consists of a pair of treatments that appear in \({0.1n}\) blocks. 
We have a \({(5, n, 0.4n, 2, 0.1n)}\)-design for short.
Finite sample performances of the AMC and the NB procedures are evaluated for controlling FWER and constructing SCIs.
We simulate data from the following standard linear mixed effect model:
\begin{equation}\label{eq:lmem}
X_{ik} = \theta_k + \beta_i + \epsilon_{ik} \textnormal{ for } i \in \mathcal{B}_k \textnormal{ and } k = 1, \dots, 5,
\end{equation}
where both \({\beta_i}\) and \({\epsilon_{ik}}\) are \({\textnormal{i.i.d.}}\) random variables for block effects and errors, respectively. 
The null hypothesis for treatment pair \({(k, l)}\) is \({H_{kl} : \theta_k - \theta_l = 0}\) for \({k, l} = {1, \dots, 5}\) with \({k} < {l}\). 
We denote the pairwise differences between treatment effects by \({\delta_j}\) for \({j} = {1, \dots, 10}\), with the corresponding hypothesis \({H_j}\) and test statistic \({T_{nj}}\). 

For comparison, we consider the single-step procedure proposed by \citet{hothorn2008simultaneous} as a benchmark. 
This procedure (henceforth HBW) is based on an asymptotic multivariate normal distribution for the point estimates and a consistent plug-in estimate of the associated covariance matrix. 
We apply HBW to restricted maximum likelihood estimates of \({\delta_j}\), assuming the additive form and compound symmetry that are present in the model.
We refer to \citet{hothorn2008simultaneous} for technical details.

We fix the level \({\alpha}\) at \({0.05}\) throughout the simulations. 
The \({\beta_i}\) and \({\epsilon_{ik}}\) are simulated from three different pairs of scenarios:
\begin{align*}
\textnormal{S1-1. } &\beta_i \sim N(0, 1) \textnormal{ and } \epsilon_{ik} \sim N(0, 1); \\
\textnormal{S1-2. } &\beta_i \sim N(0, 0.1), \epsilon_{ik} \sim N(0, 1) \textnormal{ for } k = 1,2,3,4, \textnormal{ and } \epsilon_{i5} \sim N(0, 9); \\
\textnormal{S2-1. } &\beta_i \sim \textnormal{Gamma}(2, 1) \textnormal{ and } \epsilon_{ik} \sim t(6); \\
\textnormal{S2-2. } &\beta_i \sim \textnormal{Gamma}(10, 0.1), \epsilon_{ik} \sim t(6) \textnormal{ for } k = 1,2,3,4, \textnormal{ and } \epsilon_{i5} \sim U(-5, 5); \\
\textnormal{S3-1. } &\beta_i \sim U(-0.5, 0.5) \textnormal{ and } \epsilon_{ik} \sim U(-0.5, 0.5); \\
\textnormal{S3-2. } &\beta_i \sim U(-0.1, 0.1), \epsilon_{ik} \sim U(-0.5, 0.5) \textnormal{ for } k = 1,2,3,4, \textnormal{ and } \epsilon_{i5} \sim t(3).
\end{align*}
where \({\textnormal{Gamma}(2, 1)}\) denotes a Gamma distribution with shape parameter \({2}\) and scale parameter \({1}\). 
Each pair of scenarios has a distinct distributional specification. 
In each pair, the first scenario is of the form \eqref{eq:lmem}.
The second scenario, however, has negligible block effects and larger variance for the fifth treatment, breaking some assumptions of the model.
In each scenario, we consider three different numbers of blocks \({n} \in {\{50, 100, 200\}}\) and three different values of \({\theta}\) to vary the number of true null hypotheses.
Given specific values of \({n}\) and \({\theta}\), simulation results for the AMC procedure are obtained as follows. 
For \({S} = {\num{10000}}\) simulation runs indexed by \({s}\):
\begin{step}
Simulate data from the given scenario and compute \({T_n(s)}\).
\end{step}
\begin{step}
With \({B} = {\num{10000}}\), apply the AMC procedure in Algorithm \ref{alg:AMC} to obtain \({c_\alpha^\textnormal{AMC}(s)}\), and compute the SCI \({I_{j}^{\textnormal{AMC}}(s)}\) and its length \({\vert I_j^{\textnormal{AMC}}(s)\vert}\) for each \({j}\).
\end{step}
The empirical FWER, average length (AL), and coverage probability (CP) of the SCIs are calculated as 
\begin{align*}
\textnormal{FWER} 
&= \frac{1}{S}\sum_{s = 1}^S
1\left(
\max \left \{T_{nj}(s): j \in \mathcal{I}_0 \right \} > c_\alpha^\textnormal{AMC}(s)
\right); \\
\textnormal{AL} 
&= 
\frac{1}{S}\sum_{s = 1}^S 
\left(\frac{1}{10}\sum_{j = 1}^{10}\Big\vert I_j^{\textnormal{AMC}}(s)\Big\vert \right);\\
\textnormal{CP} 
&= 
\frac{1}{S}\sum_{s = 1}^S 
1\left(
\delta_j \in I_{j}^{\textnormal{AMC}}(s) \textnormal{ for all } j
\right).
\end{align*}

The results for the NB procedure are obtained similarly. 
Step (\romannumeral 1) is modified to set up bootstrap sampling that respects \({H_0}\).
Before drawing the bootstrap replicates, pass from \({X_{ik}}\) to
\begin{equation}\label{eq:null transformation}
\widetilde{X}_{ik} = X_{ik} - \overline{X}_{k},
\end{equation}
where \({\overline{X}_{k}} = {r_k^{-1}\sum_{i \in \mathcal{B}_k}X_{ik}}\) is the maximum empirical likelihood estimate for \({\theta_k}\).
Applying \cref{alg:NB} in Step (\romannumeral 2), we obtain the same \({T_n}\) but the cutoff \({c_\alpha^\textnormal{NB}}\) is different from \({c_\alpha^\textnormal{AMC}}\), which produces different SCIs, FWER, AL, and CP.
All simulations are performed in \textsf{R} \citep{R}.
We implement AMC and NB with the \pkg{melt} package \citep{melt}.
For HBW, we fit \eqref{eq:lmem} via the \pkg{lme4} package \citep{bates2015fitting} and then pass the result to the \pkg{multcomp} package \citep{hothorn2008simultaneous}.

\cref{tab:S1,tab:S2,tab:S3} summarize the simulation results. 
In all scenarios and procedures, the FWER is largest for all \({n}\) when \({H_0}\) holds and decreases as the number of false hypotheses increases since there are fewer opportunities to reject the true null hypotheses. 
By construction, AL and CP are not related to \({\theta}\).
The intervals are shorter when the model generates less variation in the data. 
FWER and CP approach their respective targets, 0.05 and 0.95, under \({H_0}\) as \({n}\) increases. 
For AMC, FWER and CP are quite far from the targets when \({n} = {50}\) and are also sensitive to the distribution of the data. 
As can be seen from \cref{tab:S2}, the FWER and CP tend to be worse when the distribution is highly skewed and has a thick tail. The estimates computed with only 20 observations per treatment can be inaccurate in the presence of skewness and outliers. 
On the other hand, NB provides FWER and CP close to target even when \({n} = {50}\). 
NB outperforms AMC in FWER and CP but has larger AL in all scenarios. 
This finding is consistent with our experience that \({c_\alpha^\textnormal{AMC}} < {c_\alpha^\textnormal{NB}}\) holds in most cases and in keeping with the slow convergence of Wald-type statistics to chi-square distributions \citep[see, e.g.][]{pauly2015asymptotic}.
The performances of AMC and NB are similar when \({n} = {200}\) or when the data range is restricted (\cref{tab:S3}).
Interestingly, NB is more conservative when \({n} = {50}\) than when \({n} = {100}\) or \({n} = {200}\).
This is partly due to the higher chance that a bootstrap sample may not satisfy the convex hull constraint, contributing to the large cutoff of NB (see \cref{rmk:nb}).

\begin{table}
\def~{\hphantom{0}}
\tbl{Simulation results under scenario S1.
}
{
\begin{tabular}{ccccccccccc}
\multicolumn{2}{c}{} &
\multicolumn{3}{c}{AMC} &
\multicolumn{3}{c}{NB} &
\multicolumn{3}{c}{HBW}\\  
\({n}\) & \({(\theta_1, \theta_2)}\) & FWER & AL & CP (\%) & FWER & AL & CP (\%) & FWER & AL & CP (\%)\\ 
&&\multicolumn{8}{c}{S1-1}\\
\multirow{3}{*}{\({50}\)} 
& \({(0, 0)}\) 
& \({0.077}\) & \({2.284}\) & \({92.4}\) 
& \({0.043}\) & \({2.498}\) & \({95.7}\) 
& \({0.065}\) & \({1.973}\) & \({93.5}\) \\ 
& \({(1, 0)}\) 
& \({0.057}\) & \({2.283}\) & \({91.4}\) 
& \({0.034}\) & \({2.496}\) & \({94.7}\) 
& \({0.044}\) & \({1.974}\) & \({93.3}\) \\ 
& \({(2, 1)}\) 
& \({0.030}\) & \({2.282}\) & \({91.8}\) 
& \({0.018}\) & \({2.491}\) & \({95.0}\) 
& \({0.026}\) & \({1.970}\) & \({93.4}\) \\[2ex] 
\multirow{3}{*}{\({100}\)} 
& \({(0, 0)}\) 
& \({0.061}\) & \({1.628}\) & \({93.9}\) 
& \({0.051}\) & \({1.674}\) & \({94.9}\) 
& \({0.058}\) & \({1.401}\) & \({94.2}\) \\ 
& \({(1, 0)}\) 
& \({0.036}\) & \({1.629}\) & \({94.3}\) 
& \({0.030}\) & \({1.674}\) & \({95.4}\) 
& \({0.036}\) & \({1.403}\) & \({94.6}\) \\ 
& \({(2, 1)}\) 
& \({0.021}\) & \({1.628}\) & \({93.8}\) 
& \({0.017}\) & \({1.673}\) & \({94.9}\) 
& \({0.020}\) & \({1.401}\) & \({94.3}\) \\[2ex] 
\multirow{3}{*}{\({200}\)} 
& \({(0, 0)}\) 
& \({0.053}\) & \({1.148}\) & \({94.7}\) 
& \({0.049}\) & \({1.160}\) & \({95.1}\) 
& \({0.054}\) & \({0.993}\) & \({94.6}\) \\ 
& \({(1, 0)}\) 
& \({0.037}\) & \({1.149}\) & \({94.3}\) 
& \({0.033}\) & \({1.161}\) & \({94.9}\) 
& \({0.035}\) & \({0.994}\) & \({94.5}\) \\ 
& \({(2, 1)}\) 
& \({0.017}\) & \({1.149}\) & \({95.3}\) 
& \({0.017}\) & \({1.158}\) & \({95.4}\) 
& \({0.018}\) & \({0.994}\) & \({94.9}\) \\\\ 
&&\multicolumn{8}{c}{S1-2}\\
\multirow{3}{*}{\({50}\)} 
& \({(0, 0)}\) 
& \({0.078}\) & \({2.574}\) & \({92.2}\) 
& \({0.041}\) & \({2.859}\) & \({95.9}\) 
& \({0.101}\) & \({2.783}\) & \({89.9}\) \\ 
& \({(1, 0)}\) 
& \({0.058}\) & \({2.569}\) & \({91.7}\) 
& \({0.028}\) & \({2.854}\) & \({96.0}\) 
& \({0.097}\) & \({2.775}\) & \({89.1}\) \\ 
& \({(2, 1)}\) 
& \({0.031}\) & \({2.572}\) & \({91.7}\) 
& \({0.018}\) & \({2.829}\) & \({94.8}\) 
& \({0.078}\) & \({2.778}\) & \({89.5}\) \\[2ex] 
\multirow{3}{*}{\({100}\)} 
& \({(0, 0)}\) 
& \({0.061}\) & \({1.845}\) & \({94.0}\) 
& \({0.052}\) & \({1.905}\) & \({94.8}\) 
& \({0.101}\) & \({1.979}\) & \({89.9}\) \\ 
& \({(1, 0)}\) 
& \({0.041}\) & \({1.846}\) & \({93.7}\) 
& \({0.033}\) & \({1.906}\) & \({94.8}\) 
& \({0.092}\) & \({1.979}\) & \({89.7}\) \\ 
& \({(2, 1)}\) 
& \({0.021}\) & \({1.847}\) & \({93.9}\) 
& \({0.017}\) & \({1.897}\) & \({94.5}\) 
& \({0.075}\) & \({1.980}\) & \({89.7}\) \\[2ex] 
\multirow{3}{*}{\({200}\)} 
& \({(0, 0)}\) 
& \({0.053}\) & \({1.305}\) & \({94.7}\) 
& \({0.051}\) & \({1.320}\) & \({95.0}\) 
& \({0.094}\) & \({1.405}\) & \({90.6}\) \\ 
& \({(1, 0)}\) 
& \({0.036}\) & \({1.304}\) & \({94.4}\) 
& \({0.035}\) & \({1.320}\) & \({94.8}\) 
& \({0.084}\) & \({1.404}\) & \({90.7}\) \\ 
& \({(2, 1)}\) 
& \({0.020}\) & \({1.305}\) & \({94.4}\) 
& \({0.019}\) & \({1.307}\) & \({94.5}\) 
& \({0.071}\) & \({1.405}\) & \({90.4}\)
\end{tabular}
}
\label{tab:S1}
\begin{tabnote}
\({\theta_3}\), \({\theta_4}\), and \({\theta_5}\) are set to zero. The largest standard error of the results is \({0.003}\) when \({n} = {50}\).
\end{tabnote}
\end{table}

\begin{table}
\def~{\hphantom{0}}
\tbl{Simulation results under scenario S2.
}
{
\begin{tabular}{ccccccccccc}
\multicolumn{2}{c}{} &
\multicolumn{3}{c}{AMC} &
\multicolumn{3}{c}{NB} &
\multicolumn{3}{c}{HBW}\\  
\({n}\) & \({(\theta_1, \theta_2)}\) & FWER & AL & CP (\%) & FWER & AL & CP (\%) & FWER & AL & CP (\%)\\ 
&&\multicolumn{8}{c}{S2-1}\\
\multirow{3}{*}{\({50}\)} 
& \({(0, 0)}\) 
& \({0.091}\) & \({3.032}\) & \({90.1}\) 
& \({0.047}\) & \({3.435}\) & \({95.2}\) 
& \({0.064}\) & \({2.444}\) & \({93.6}\) \\ 
& \({(1, 0)}\) 
& \({0.064}\) & \({3.035}\) & \({90.2}\) 
& \({0.031}\) & \({3.439}\) & \({95.1}\) 
& \({0.044}\) & \({2.445}\) & \({93.3}\) \\ 
& \({(2, 1)}\) 
& \({0.032}\) & \({3.029}\) & \({90.9}\)
& \({0.015}\) & \({3.431}\) & \({95.1}\) 
& \({0.023}\) & \({2.439}\) & \({93.6}\) \\[2ex] 
\multirow{3}{*}{\({100}\)} 
& \({(0, 0)}\) 
& \({0.072}\) & \({2.167}\) & \({92.8}\) 
& \({0.053}\) & \({2.270}\) & \({94.7}\) 
& \({0.057}\) & \({1.739}\) & \({94.3}\) \\ 
& \({(1, 0)}\) 
& \({0.041}\) & \({2.168}\) & \({93.4}\) 
& \({0.031}\) & \({2.271}\) & \({95.1}\) 
& \({0.036}\) & \({1.741}\) & \({94.3}\) \\ 
& \({(2, 1)}\) 
& \({0.023}\) & \({2.171}\) & \({93.2}\) 
& \({0.017}\) & \({2.275}\) & \({95.1}\) 
& \({0.021}\) & \({1.742}\) & \({94.6}\) \\[2ex] 
\multirow{3}{*}{\({200}\)} 
& \({(0, 0)}\) 
& \({0.056}\) & \({1.524}\) & \({94.5}\) 
& \({0.049}\) & \({1.552}\) & \({95.1}\) 
& \({0.053}\) & \({1.234}\) & \({94.7}\) \\ 
& \({(1, 0)}\) 
& \({0.039}\) & \({1.526}\) & \({94.0}\) 
& \({0.035}\) & \({1.553}\) & \({94.7}\) 
& \({0.035}\) & \({1.234}\) & \({94.7}\) \\ 
& \({(2, 1)}\) 
& \({0.020}\) & \({1.526}\) & \({94.2}\) 
& \({0.018}\) & \({1.554}\) & \({94.7}\) 
& \({0.019}\) & \({1.236}\) & \({94.4}\) \\\\ 
&&\multicolumn{8}{c}{S2-2}\\
\multirow{3}{*}{\({50}\)} 
& \({(0, 0)}\) 
& \({0.092}\) & \({2.761}\) & \({90.8}\) 
& \({0.047}\) & \({3.091}\) & \({95.3}\) 
& \({0.093}\) & \({2.917}\) & \({90.7}\) \\ 
& \({(1, 0)}\) 
& \({0.055}\) & \({2.765}\) & \({91.2}\) 
& \({0.027}\) & \({3.097}\) & \({95.6}\) 
& \({0.082}\) & \({2.922}\) & \({90.4}\) \\ 
& \({(2, 1)}\) 
& \({0.029}\) & \({2.765}\) & \({90.7}\) 
& \({0.014}\) & \({3.079}\) & \({94.8}\) 
& \({0.069}\) & \({2.923}\) & \({90.3}\) \\[2ex] 
\multirow{3}{*}{\({100}\)} 
& \({(0, 0)}\) 
& \({0.066}\) & \({1.992}\) & \({93.4}\) 
& \({0.051}\) & \({2.074}\) & \({94.9}\) 
& \({0.089}\) & \({2.077}\) & \({91.1}\) \\ 
& \({(1, 0)}\) 
& \({0.040}\) & \({1.991}\) & \({93.7}\) 
& \({0.031}\) & \({2.072}\) & \({95.1}\) 
& \({0.078}\) & \({2.075}\) & \({91.1}\) \\ 
& \({(2, 1)}\) 
& \({0.022}\) & \({1.993}\) & \({93.5}\) 
& \({0.017}\) & \({2.072}\) & \({94.8}\) 
& \({0.064}\) & \({2.077}\) & \({90.9}\) \\[2ex] 
\multirow{3}{*}{\({200}\)} 
& \({(0, 0)}\) 
& \({0.056}\) & \({1.412}\) & \({94.4}\)
& \({0.050}\) & \({1.435}\) & \({95.1}\) 
& \({0.086}\) & \({1.472}\) & \({91.5}\) \\ 
& \({(1, 0)}\) 
& \({0.036}\) & \({1.414}\) & \({94.6}\) 
& \({0.032}\) & \({1.438}\) & \({95.2}\) 
& \({0.072}\) & \({1.474}\) & \({91.9}\) \\ 
& \({(2, 1)}\) 
& \({0.019}\) & \({1.415}\) & \({94.4}\) 
& \({0.017}\) & \({1.432}\) & \({94.8}\) 
& \({0.059}\) & \({1.474}\) & \({91.6}\)
\end{tabular}
}
\label{tab:S2}
\begin{tabnote}
\({\theta_3}\), \({\theta_4}\), and \({\theta_5}\) are set to zero. 
The largest standard error of the results is \({0.003}\) when \({n} = {50}\).
\end{tabnote}
\end{table}

\begin{table}
\def~{\hphantom{0}}
\tbl{Simulation results under scenario S3.}
{
\begin{tabular}{ccccccccccc}
\multicolumn{2}{c}{} &
\multicolumn{3}{c}{AMC} &
\multicolumn{3}{c}{NB} &
\multicolumn{3}{c}{HBW}\\  
\({n}\) & \({(\theta_1, \theta_2)}\) & FWER & AL & CP (\%) & FWER & AL & CP (\%) & FWER & AL & CP (\%)\\ 
&&\multicolumn{8}{c}{S3-1}\\
\multirow{3}{*}{\({50}\)} 
& \({(0, 0)}\) 
& \({0.078}\) & \({0.652}\) & \({92.3}\) 
& \({0.046}\) & \({0.699}\) & \({95.4}\) 
& \({0.070}\) & \({0.571}\) & \({93.1}\) \\ 
& \({(1/8, 0)}\) 
& \({0.054}\) & \({0.652}\) & \({92.2}\) 
& \({0.034}\) & \({0.699}\) & \({94.9}\) 
& \({0.046}\) & \({0.571}\) & \({93.3}\) \\ 
& \({(1/4, 1/8)}\) 
& \({0.029}\) & \({0.651}\) & \({92.2}\) 
& \({0.018}\) & \({0.699}\) & \({94.9}\) 
& \({0.024}\) & \({0.571}\) & \({93.3}\) \\[2ex] 
\multirow{3}{*}{\({100}\)} 
& \({(0, 0)}\) 
& \({0.060}\) & \({0.465}\) & \({94.0}\) 
& \({0.051}\) & \({0.474}\) & \({94.9}\) 
& \({0.058}\) & \({0.405}\) & \({94.2}\) \\ 
& \({(1/8, 0)}\) 
& \({0.039}\) & \({0.465}\) & \({93.8}\) 
& \({0.035}\) & \({0.474}\) & \({94.4}\) 
& \({0.041}\) & \({0.405}\) & \({93.7}\) \\ 
& \({(1/4, 1/8)}\) 
& \({0.021}\) & \({0.465}\) & \({94.2}\) 
& \({0.019}\) & \({0.475}\) & \({95.0}\) 
& \({0.021}\) & \({0.406}\) & \({94.4}\) \\[2ex] 
\multirow{3}{*}{\({200}\)} 
& \({(0, 0)}\) 
& \({0.053}\) & \({0.329}\) & \({94.8}\) 
& \({0.050}\) & \({0.332}\) & \({95.0}\) 
& \({0.055}\) & \({0.287}\) & \({94.5}\) \\ 
& \({(1/8, 0)}\)    
& \({0.034}\) & \({0.329}\) & \({94.9}\) 
& \({0.032}\) & \({0.332}\) & \({95.1}\) 
& \({0.033}\) & \({0.287}\) & \({94.9}\) \\ 
& \({(1/4, 1/8)}\) 
& \({0.018}\) & \({0.329}\) & \({94.9}\) 
& \({0.017}\) & \({0.332}\) & \({95.3}\) 
& \({0.017}\) & \({0.287}\) & \({94.6}\) \\\\ 
&&\multicolumn{8}{c}{S3-2}\\
\multirow{3}{*}{\({50}\)} 
& \({(0, 0)}\) 
& \({0.108}\) & \({1.062}\) & \({90.9}\) 
& \({0.033}\) & \({1.536}\) & \({96.6}\) 
& \({0.117}\) & \({1.277}\) & \({88.3}\) \\ 
& \({(1/8, 0)}\) 
& \({0.079}\) & \({1.077}\) & \({91.1}\) 
& \({0.024}\) & \({1.567}\) & \({96.6}\) 
& \({0.109}\) & \({1.304}\) & \({88.3}\) \\ 
& \({(1/4, 1/8)}\) 
& \({0.056}\) & \({1.071}\) & \({91.2}\) 
& \({0.015}\) & \({1.547}\) & \({96.7}\) 
& \({0.097}\) & \({1.292}\) & \({88.5}\) \\[2ex] 
\multirow{3}{*}{\({100}\)} 
& \({(0, 0)}\) 
& \({0.070}\) & \({0.807}\) & \({93.3}\) 
& \({0.045}\) & \({0.886}\) & \({95.6}\) 
& \({0.105}\) & \({0.940}\) & \({89.5}\) \\ 
& \({(1/8, 0)}\) 
& \({0.049}\) & \({0.803}\) & \({93.2}\) 
& \({0.032}\) & \({0.877}\) & \({95.3}\) 
& \({0.096}\) & \({0.934}\) & \({89.7}\) \\ 
& \({(1/4, 1/8)}\) 
& \({0.032}\) & \({0.809}\) & \({93.2}\) 
& \({0.022}\) & \({0.886}\) & \({95.1}\) 
& \({0.087}\) & \({0.943}\) & \({89.7}\) \\[2ex] 
\multirow{3}{*}{\({200}\)} 
& \({(0, 0)}\) 
& \({0.061}\) & \({0.580}\) & \({93.9}\) 
& \({0.051}\) & \({0.603}\) & \({94.9}\) 
& \({0.106}\) & \({0.673}\) & \({89.4}\) \\ 
& \({(1/8, 0)}\) 
& \({0.041}\) & \({0.583}\) & \({94.0}\) 
& \({0.035}\) & \({0.608}\) & \({94.9}\) 
& \({0.101}\) & \({0.678}\) & \({89.3}\) \\ 
& \({(1/4, 1/8)}\) 
& \({0.025}\) & \({0.583}\) & \({94.2}\) 
& \({0.020}\) & \({0.607}\) & \({95.3}\) 
& \({0.092}\) & \({0.678}\) & \({89.3}\)
\end{tabular}
}
\label{tab:S3}
\begin{tabnote}
\({\theta_3}\), \({\theta_4}\), and \({\theta_5}\) are set to zero. The largest standard error of the results is \({0.007}\) when \({n} = {50}\).
\end{tabnote}
\end{table}

\begin{figure}[!t]
\centering
\includegraphics[width=\textwidth]{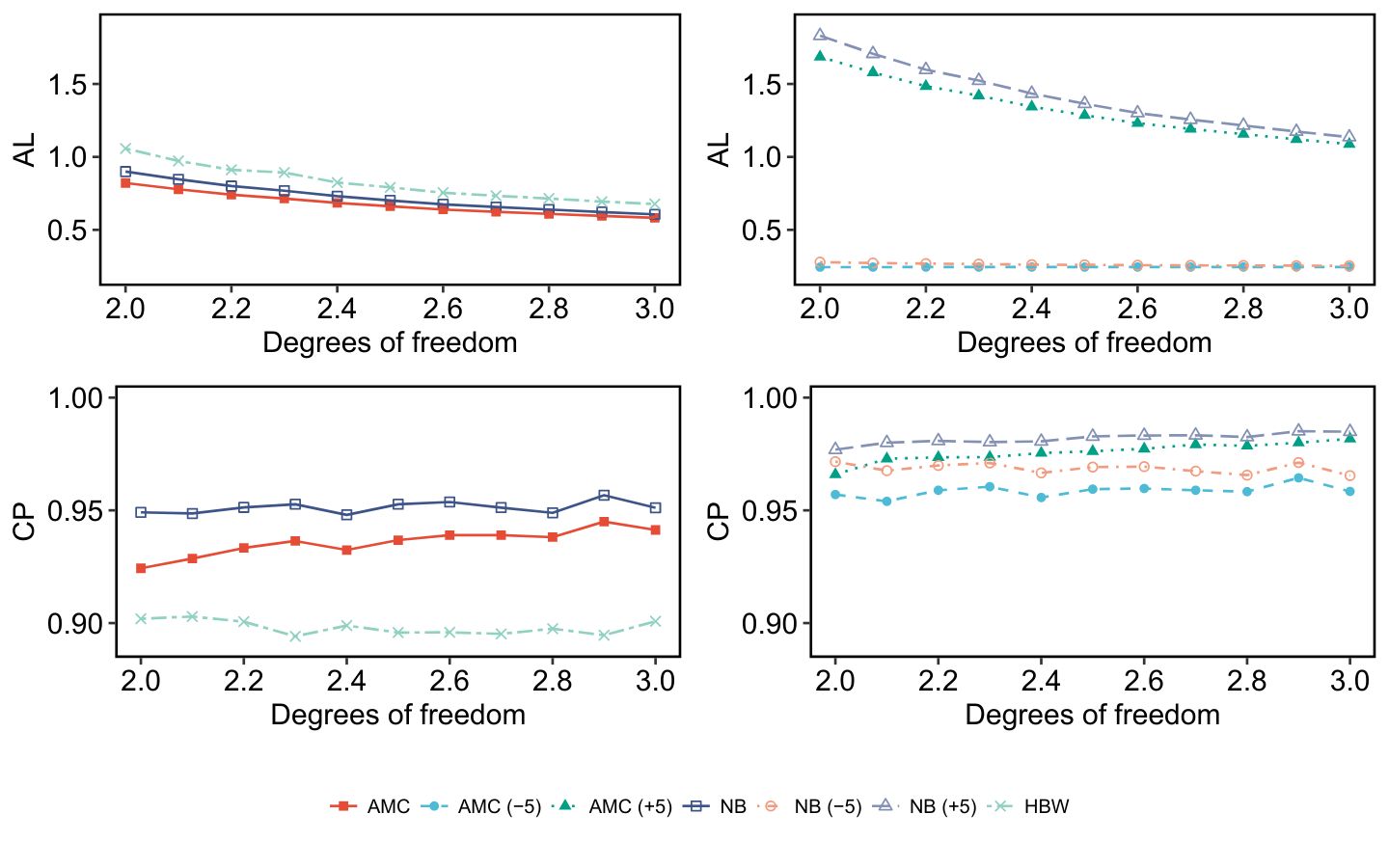}
\caption{
AL and CP of S3-2 with varying degrees of freedom for the distribution of \({\epsilon_{i5}}\) when \({n} = {200}\) and \({\theta} = {(0, 0, 0, 0, 0)}\).
For AMC and NB, separate results are presented for the comparisons that involve \({\theta_5}\) (\({+5}\)) and those that do not (\({-5}\)).
AL and CP are computed group-wise.
}
\label{fig:df}
\end{figure}

The HBW procedure, contrary to the empirical likelihood-based procedures, depends heavily on the assumptions in \eqref{eq:lmem}. 
HBW performs well in scenarios where the compound symmetry assumption is met (S1-1, S2-1, and S3-1). 
FWER and CP are robust across different distributions for the block effects and the errors. 
Except for \({n} = {200}\), HBW outperforms AMC and comes close to NB with considerably shorter AL. 
In scenarios where compound symmetry is violated (S1-2, S2-2, and S3-2), however, HBW shows a substantial performance deterioration. 
The AL of HBW is larger than those of AMC and NB when \({n}\) is \({100}\) or \({200}\). 
FWER and CP are far from their target values, and the rate at which they improve is much slower. 
\cref{fig:df} further shows the impact of the violation on AL and CP by gradually decreasing the degrees of freedom for the distribution of \({\epsilon_{i5}}\) in S3-2 when \({n} = {200}\) and \({\theta} = {(0, 0, 0, 0, 0)}\).
The AL of AMC and NB is much larger for the intervals with \({\theta_5}\) than the rest, and only the AL of these intervals increases as the degrees of freedom decrease to \({2}\) (infinite variance).
As a result of this adjustment, the CP of the individual interval is maintained above \({0.95}\) for AMC and NB.
In contrast, all intervals of HBW have the same length.
This implies that the intervals with \({\theta_5}\) are not wide enough as SCIs, causing the under-coverage shown in \cref{fig:df}.
For AMC and NB, additional simulation results for gFWER control are also shown in \cref{tab:gfwer}.

\begin{table}
\def~{\hphantom{0}}
\tbl{
Simulation results for the gFWER control. 
For the three simulation scenarios, gFWER of each procedure is computed below for \({v} = {2, 3, 4, 5}\) with \({\theta} = (0, 0, 0, 0, 0)\).
}
{
\begin{tabular}{ccccccccc}
\multicolumn{1}{c}{} &
\multicolumn{4}{c}{AMC} &
\multicolumn{4}{c}{NB} \\  
\({n}\)
& \({v} = {2}\) & \({v} = {3}\) & \({v} = {4}\) & \({v} = {5}\) 
& \({v} = {2}\) & \({v} = {3}\) & \({v} = {4}\) & \({v} = {5}\) \\
& \multicolumn{8}{c}{S1-1}  \\
\({50}\) 
& \({0.065}\)
& \({0.057}\) 
& \({0.052}\) 
& \({0.054}\) 
& \({0.041}\) 
& \({0.039}\) 
& \({0.038}\) 
& \({0.040}\) \\
\({100}\)
& \({0.055}\) 
& \({0.054}\) 
& \({0.054}\) 
& \({0.054}\) 
& \({0.048}\) 
& \({0.049}\) 
& \({0.048}\) 
& \({0.046}\) \\
\({200}\)
& \({0.056}\) 
& \({0.054}\) 
& \({0.049}\) 
& \({0.053}\) 
& \({0.053}\) 
& \({0.050}\) 
& \({0.046}\) 
& \({0.050}\) \\[2ex]
& \multicolumn{8}{c}{S1-2}  \\
\({50}\) 
& \({0.067}\) 
& \({0.058}\) 
& \({0.057}\) 
& \({0.058}\) 
& \({0.042}\) 
& \({0.040}\) 
& \({0.042}\) 
& \({0.043}\) \\
\({100}\)
& \({0.056}\) 
& \({0.057}\) 
& \({0.056}\) 
& \({0.057}\) 
& \({0.048}\) 
& \({0.051}\) 
& \({0.051}\) 
& \({0.048}\) \\
\({200}\)
& \({0.055}\) 
& \({0.051}\) 
& \({0.051}\) 
& \({0.051}\) 
& \({0.051}\) 
& \({0.048}\) 
& \({0.048}\) 
& \({0.048}\) \\[2ex]
& \multicolumn{8}{c}{S2-1}  \\
\({50}\) 
& \({0.074}\) 
& \({0.064}\) 
& \({0.058}\) 
& \({0.058}\) 
& \({0.044}\) 
& \({0.043}\) 
& \({0.040}\) 
& \({0.041}\) \\
\({100}\)
& \({0.065}\) 
& \({0.063}\) 
& \({0.058}\) 
& \({0.056}\) 
& \({0.054}\) 
& \({0.050}\) 
& \({0.050}\) 
& \({0.046}\) \\
\({200}\)
& \({0.055}\) 
& \({0.053}\) 
& \({0.050}\) 
& \({0.051}\) 
& \({0.051}\) 
& \({0.049}\) 
& \({0.047}\) 
& \({0.046}\) \\[2ex]
& \multicolumn{8}{c}{S2-2}  \\
\({50}\) 
& \({0.076}\) 
& \({0.067}\) 
& \({0.062}\) 
& \({0.060}\) 
& \({0.044}\) 
& \({0.045}\) 
& \({0.044}\) 
& \({0.043}\) \\
\({100}\)
& \({0.061}\) 
& \({0.055}\) 
& \({0.057}\) 
& \({0.058}\) 
& \({0.051}\) 
& \({0.048}\) 
& \({0.050}\) 
& \({0.050}\) \\
\({200}\)
& \({0.055}\) 
& \({0.053}\) 
& \({0.051}\) 
& \({0.049}\) 
& \({0.050}\) 
& \({0.049}\) 
& \({0.049}\) 
& \({0.046}\) \\[2ex]
& \multicolumn{8}{c}{S3-1}  \\
\({50}\) 
& \({0.067}\) 
& \({0.064}\) 
& \({0.061}\) 
& \({0.056}\) 
& \({0.049}\) 
& \({0.049}\) 
& \({0.047}\) 
& \({0.044}\) \\
\({100}\)
& \({0.053}\) 
& \({0.051}\) 
& \({0.051}\) 
& \({0.055}\) 
& \({0.046}\) 
& \({0.047}\) 
& \({0.046}\) 
& \({0.048}\) \\
\({200}\)
& \({0.052}\) 
& \({0.047}\) 
& \({0.051}\) 
& \({0.054}\) 
& \({0.049}\) 
& \({0.046}\) 
& \({0.048}\) 
& \({0.050}\) \\[2ex]
& \multicolumn{8}{c}{S3-2}  \\
\({50}\) 
& \({0.097}\) 
& \({0.091}\) 
& \({0.087}\) 
& \({0.071}\) 
& \({0.039}\) 
& \({0.044}\) 
& \({0.047}\) 
& \({0.046}\) \\
\({100}\)
& \({0.070}\) 
& \({0.072}\) 
& \({0.073}\) 
& \({0.063}\) 
& \({0.048}\) 
& \({0.050}\) 
& \({0.051}\) 
& \({0.048}\) \\
\({200}\)
& \({0.062}\) 
& \({0.060}\) 
& \({0.061}\) 
& \({0.058}\) 
& \({0.049}\) 
& \({0.046}\) 
& \({0.047}\) 
& \({0.049}\) \\[2ex]
\end{tabular}
}
\label{tab:gfwer}
\begin{tabnote}
The largest standard error of the results is \({0.003}\) when \({n} = {50}\).
\end{tabnote}
\end{table}

In summary, AMC and NB show robust performance without relying on restrictive model assumptions. 
Notably, NB successfully approximates the target error rate and CP even with small sample sizes. 
The performance gap between AMC and NB is modest for larger \({n}\), where the computational burden of the bootstrap and optimization involved in NB gives the advantage to AMC. 

\section{Application to pesticide concentration experiments}\label{sec:application}
We apply the methodology developed in \cref{sec:theory,sec:procedure} to analyze the clothianidin concentration data in \citet{alford2017translocation}. 
Clothianidin, a neonicotinoid pesticide, is a potent agonist of the nicotinic acetylcholine receptor in insects and is extensively applied in the United States to maize seeds before planting.
Quantifying the amount of clothianidin translocated into plant tissue, coupled with its potential for environmental accumulation via runoff or leaching, provides information on the costs and benefits of this delivery method. 

\citet{alford2017translocation} investigated clothianidin concentration in three regions (seed, root, and shoot) of maize plants in the early growing season. 
Two field experiments were conducted in 2014 and 2015 with four seed treatments: untreated (Naked), fungicide only (Fungicide), low rate of 0.25 mg clothianidin/kernel (Low), and high rate of 1.25 mg clothianidin/kernel (High). 
In a randomized complete block design with four blocks for each year, the treatments were applied to four plots in each block. 
Clothianidin contamination of the untreated plots was expected due to subsurface flow and proximity between plots.
Sampling was carried out at 6, 8, 10, 13, 15, 17, 20, and 34 days post-planting in 2014, and at 5, 7, 9, 12, 14, 16, 19, 47, and 61 days post-planting in 2015. 
On each sampling day, up to ten plants were randomly sampled from each plot, and three or five of them were processed for chemical analysis. 
Some plant observations were lost before the analysis. 
Root and shoot regions of the remaining observations were scored as ``complete'' (\({> 80\%}\) present) or ``incomplete'' (\({< 80\%}\) present). 
In this way, the experimental design had a hierarchical structure of sampling (year/block/days post-planting/plot) that is unbalanced and incomplete in several layers. 
The clothianidin concentration data were log-transformed to conform more closely to a normal distribution. 
\citet{alford2017translocation} fit a linear mixed model to test two contrasts: Naked \({+}\) Fungicide vs.~Low and Naked \({+}\) Fungicide vs.~High. 
\citet{jensen2018experimental} used the same data and performed a variety of post-hoc pairwise comparisons with various linear mixed models.

\begin{figure}[!t]
\captionsetup[subfigure]{justification=centering}
\centering
\begin{subfigure}{0.49\textwidth}
\includegraphics[width=\textwidth]{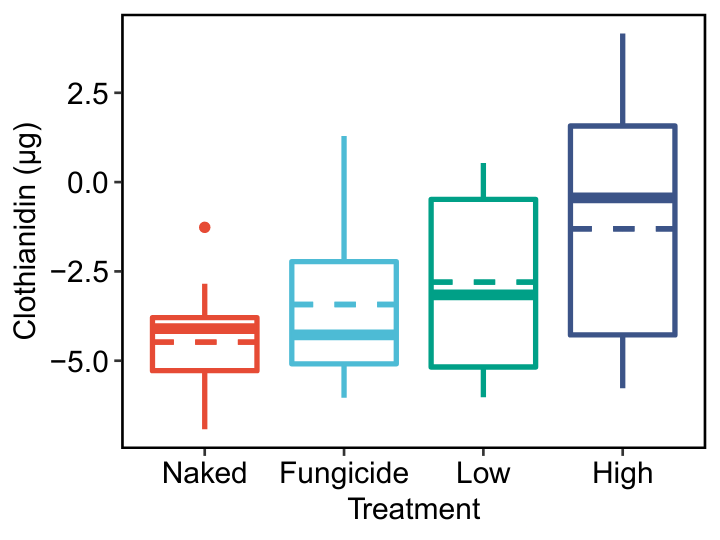}
\captionsetup{font=small}
\caption{}
\end{subfigure}
\hfill
\begin{subfigure}{0.49\textwidth}
\includegraphics[width=\textwidth]{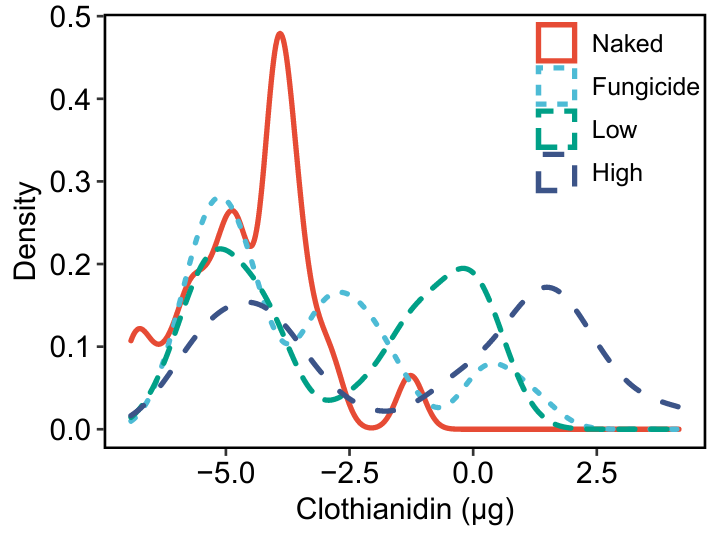}
\caption{}
\end{subfigure}
\caption{Summary of data for each treatment: (a) box plot of log transformed clothianidin concentration with median (solid line) and mean (dashed line); (b) density plot of clothianidin concentration.}
\label{fig:marginal}
\end{figure}

\begin{figure}[!t]
\centering
\includegraphics[width=\textwidth]{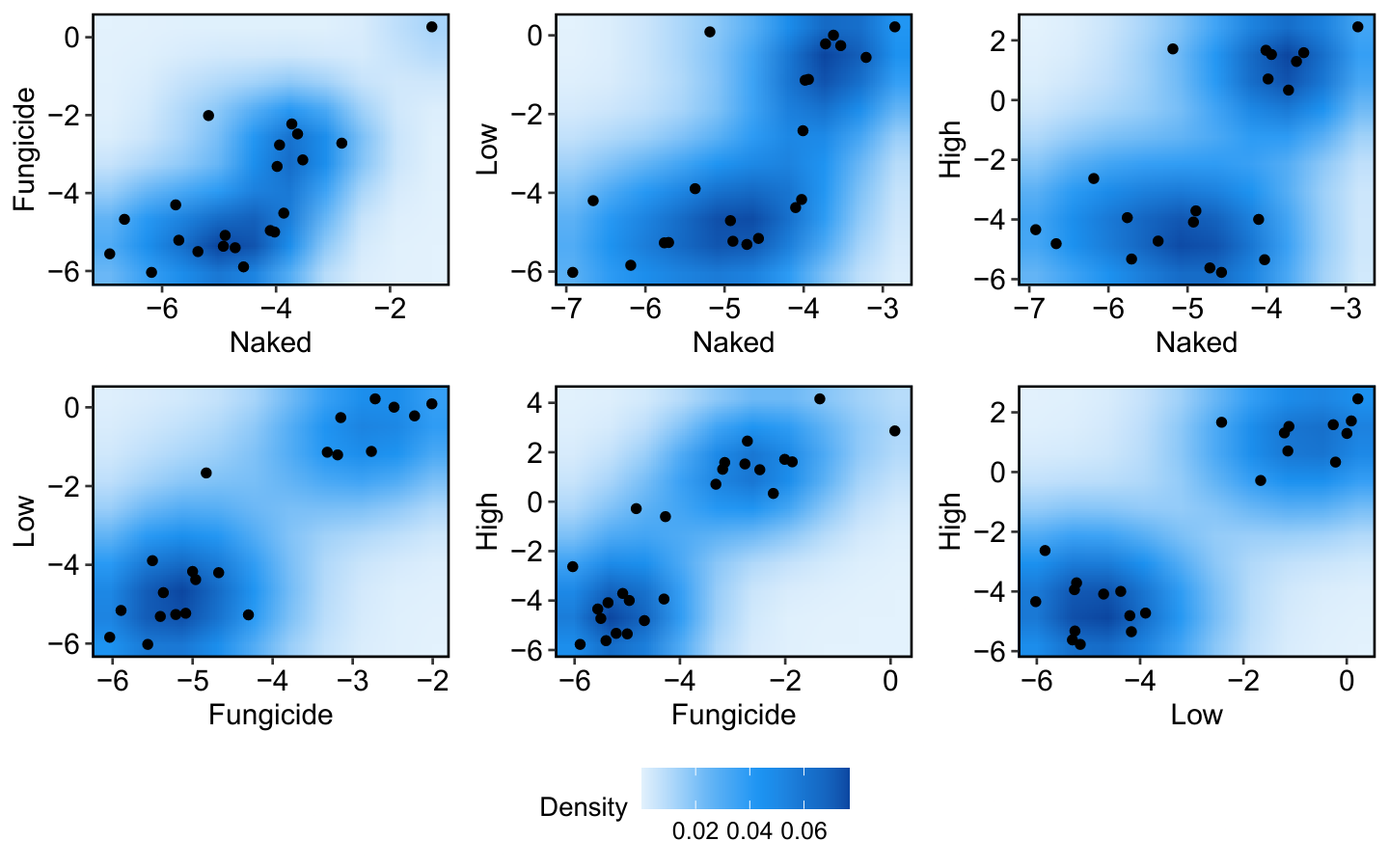}
\caption{
Pairwise scatter plots of observations. 
Each dot represents a pair of observations in a block; incomplete pairs are discarded in each plot. 
The overlaid heat maps show densities.
Many dots are located either in the bottom left or upper right corners.}
\label{fig:pairwise}
\end{figure}

We subdivide the original blocks into 68 new blocks according to the days post-planting, considering that plant tissue clothianidin dissipates over time following an exponential decay pattern \citep[Fig. 2]{alford2017translocation}. 
For illustration, we analyze only the ``incomplete'' shoot region observations. 
This results in 32 blocks. 
Plot level replicates, if any, are averaged over the treatments within these blocks, resulting in 102 observations in total. 
The decay pattern gives rise to skewed or multimodal marginal distributions for the treatments with different variances, as shown in \cref{fig:marginal}.
Regardless of the treatments, the clothianidin concentration is close to zero roughly 20 days post-planting.
Furthermore, the pairwise plots in \cref{fig:pairwise} indicate that no pairs of treatments follow a bivariate normal distribution.
Salient features of the data are heteroscedasticity, non-normality, and a violation of block-treatment additivity.

\begin{figure}[!t]
\centering
\includegraphics[width=\textwidth]{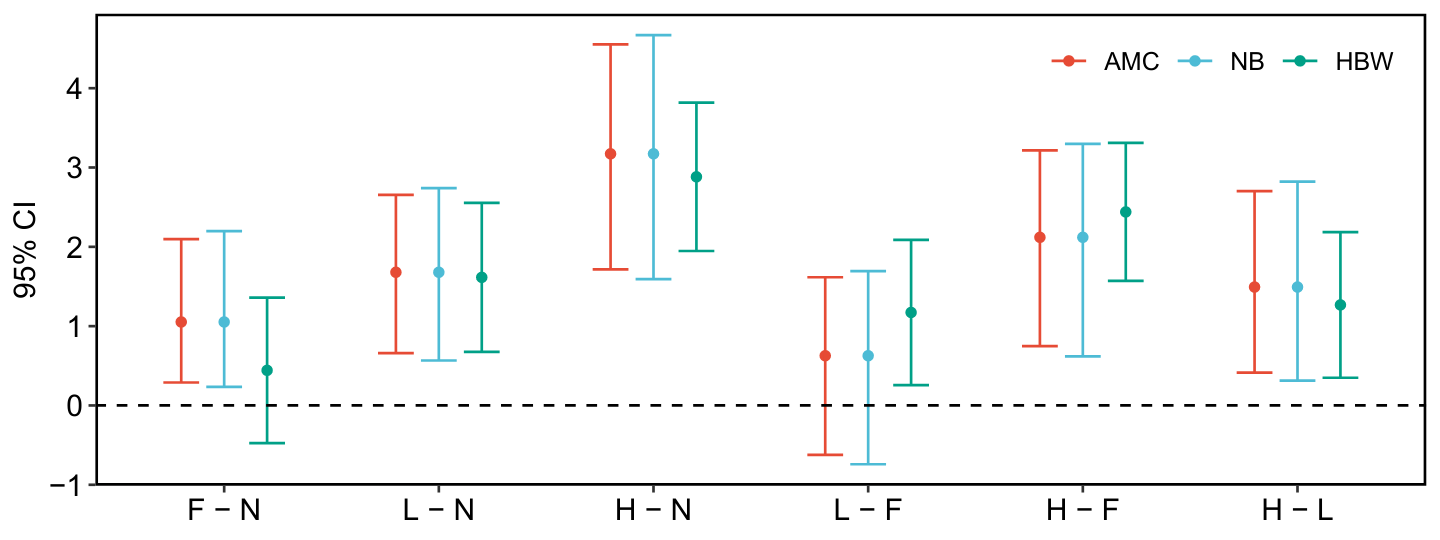}
\caption{Asymptotic \({95\%}\) simultaneous confidence intervals for pairwise comparisons. 
The estimates are given as dots inside the error bars. 
As a result of the larger cutoff, the NB interval contains the corresponding AMC interval for all comparisons.}
\label{fig:confidence intervals}
\end{figure}

We follow the procedures outlined in \cref{sec:simulation} to perform pairwise comparisons. 
We obtain estimates from the empirical likelihood and linear mixed model in \eqref{eq:lmem}; 
then, adjusted \({p}\)-values and confidence intervals are constructed using AMC, NB, and HBW.
\cref{tab:estimates} reports the estimates and \({p}\)-values, where the treatment effects are denoted by \({\theta_\textnormal{N}}\), \({\theta_\textnormal{F}}\), \({\theta_\textnormal{L}}\), and \({\theta_\textnormal{H}}\).
\begin{table}
\def~{\hphantom{0}}
\tbl{Pairwise comparisons between treatments: Naked(N), Fungicide(F), Low(L), and High(H). The estimates are obtained from empirical likelihood (EL) and the mixed effect model.}
{
\begin{tabular}{cccccc}
&
\multicolumn{2}{c}{Estimate} &
\multicolumn{3}{c}{\({p}\)-value}\\ 
Comparison & EL & Mixed model & AMC & NB & HBW\\ 
\({\theta_\textnormal{F} - \theta_\textnormal{N}}\)
& \({1.053}\) & \({0.443}\) 
& \({0.001}\) & \({0.008}\) & \({0.601}\) \\
\({\theta_\textnormal{L} - \theta_\textnormal{N}}\) 
& \({1.679}\) & \({1.615}\) 
& \({<0.001}\) & \({0.005}\) & \({<0.001}\) \\
\({\theta_\textnormal{H} - \theta_\textnormal{N}}\) 
& \({3.173}\) & \({2.883}\)
& \({<0.001}\) & \({0.001}\) & \({<0.001}\) \\
\({\theta_\textnormal{L} - \theta_\textnormal{F}}\)
& \({0.627}\) & \({1.172}\)
& \({0.503}\) & \({0.532}\) & \({0.006}\) \\
\({\theta_\textnormal{H} - \theta_\textnormal{F}}\)
& \({2.120}\) & \({2.434}\) 
& \({0.001}\) & \({0.007}\) & \({<0.001}\) \\
\({\theta_\textnormal{H} - \theta_\textnormal{L}}\)
& \({1.493}\) & \({1.268}\) 
& \({0.003}\) & \({0.015}\) & \({0.002}\)
\end{tabular}
}
\label{tab:estimates}
\begin{tabnote}
The \({p}\)-values are based on \({\num{10000}}\) Monte Carlo samples for AMC and \({\num{10000}}\) bootstrap replicates for NB.
\end{tabnote}
\end{table}
Despite the small number of blocks (\({n} = {32}\)), AMC and NB reach similar conclusions: 
the clothianidin concentration does not differ significantly between the fungicide and low rate treated plants (\({\theta_\textnormal{L} - \theta_\textnormal{F}}\)), but significant differences are observed in all the other comparisons. 
In \cref{fig:confidence intervals}, the lengths of the SCIs are similar for the two procedures, with AMC intervals being slightly shorter.
The intervals are wider for those comparisons involving some clothianidin treatment.

HBW produces a different conclusion. 
The violations of the assumptions for HBW lead to different estimates for treatment effects than under empirical likelihood (except for \({\theta_\textnormal{F}}\)). 
This leads to substantially different estimates for comparisons \({\theta_\textnormal{F} - \theta_\textnormal{N}}\) and \({\theta_\textnormal{L} - \theta_\textnormal{F}}\). 
The equal variance assumption distorts the standard errors for the comparisons. 
This contributes to the large \({p}\)-value for \({\theta_\textnormal{F} - \theta_\textnormal{N}}\) produced by HBW.

To summarize, empirical likelihood has the advantage of avoiding clearly inappropriate assumptions. 
For these data, imposing these assumptions and performing a standard analysis leads to different conclusions. 

\section{Discussion}\label{sec:discussion}
Several extensions remain open for future research. 
We are primarily interested in producing common cutoffs for pairwise comparisons and SCIs, but the AMC and NB procedures can be modified to yield common quantiles. 
Common quantile procedures are appropriate when the asymptotic multivariate chi-square distribution in \cref{thm:mvchisq} has different degrees of freedom for each marginal. 
While preserving the gFWER control, improvement in power can be achieved by adapting to stepwise procedures in both approaches.

As previously mentioned in \cref{subsec:mvchisq}, it is also possible to study other parameters and estimating functions in a similar fashion.
AMC and NB would need minor adjustment once an asymptotic multivariate distribution is established, though it can be challenging to specify the null transformation for NB. 
Due to nonconvexity, a major challenge would be the computation of the statistics in \eqref{eq:EL statistic}.
See \citet{tang2014nested} for general strategies for computing constrained empirical likelihood problems.

Finally, another interesting topic concerning multiple testing is the use of high-dimensional estimating functions with a growing number of parameters.
\citet{hjort2009extending} and \citet{tang2010penalized} investigated the feasibility of empirical likelihood methods when \({p}\), the dimension of the parameter, is allowed to increase with \({n}\).
In such high-dimensional settings, however, the typical objective is often to control other types of error rates, such as the false discovery rate, which is outside the scope of this article.

\section*{Acknowledgments}
We thank the associate editor and three reviewers for their valuable feedback and suggestions, which greatly improved the quality of this paper. 

\section*{Disclosure statement}
No potential conflict of interest was reported by the authors.

\section*{Funding}
This work was supported by the US National Science Foundation under Grants No.~SES-1921523 and DMS-2015552. 

\bibliographystyle{gNST}
\bibliography{references}

\section*{Appendix}\label{sec:appendix}
\begin{proof}[Proof of \cref{thm:mvchisq}]
By \cref{condition:consistent covariance estimation}, \({S_n(\theta_0)} \to {V}\) in probability and \({S_n(\theta_0)}\) has full rank with high probability for large \({n}\). 
Then, Proposition 1 of \citet{chaudhuri2017hamiltonian} applies and there exists an open ball around \({\theta_0}\) where \({l_n(\theta)}\) is defined. 
Adjusting \({\mathcal{N}}\) if necessary, it follows from \cref{condition:convex hull} that 
\begin{equation}\label{eq:convex hull constraint satisfied}
P\left\{0 \in \textnormal{Conv}_n(\theta) \textnormal{ for all } \theta \in \mathcal{N}\right\} \to 1.
\end{equation}
Moreover, the implicit function theorem implies that \({l_n(\theta)}\) is continuously differentiable on \({\mathcal{N}}\).

Consider any consistent estimator \({\widehat{\theta}}\) of \({\theta_0}\) such that \({\widehat{\theta} - \theta_0} = {O_{P}(a_n^{-1})}\).
From \eqref{eq:convex hull constraint satisfied}, we assume that the convex hull constraint is satisfied for \({\widehat{\theta}}\) throughout the proof. 
Let \({\widehat{g}_i} = {g(X_i, \widehat{\theta})}\) and \({Z_n} = {\max_{1 \leq i \leq n} \vert\widehat{g}_i\vert}\).
Following standard arguments as in \citet[pp.~219--222]{owen2001empirical}, write \({\widehat{\lambda}} \equiv {\widehat{\lambda}(\widehat{\theta})} = {\vert{\widehat{\lambda}}\vert\widehat{\mu}}\) for \({\vert\widehat{\mu}\vert} = {1}\), where \({\widehat{\lambda}}\) solves
\begin{equation}\label{eq:lambda solution}
\frac{1}{n}\sum_{i = 1}^n \frac{\widehat{g}_i}{1 + \widehat{\lambda}^\top \widehat{g}_i} = 0.
\end{equation}
Substituting \({1 / (1 + \widehat{\lambda}^\top\widehat{g}_i)} = {1 - \widehat{\lambda}^\top \widehat{g}_i/(1 + \widehat{\lambda}^\top\widehat{g}_i)}\) into \eqref{eq:lambda solution}, we obtain
\begin{equation*}
0 = 
\widehat{\mu}^\top G_n(\widehat{\theta}) - 
\vert\widehat{\lambda}\vert \widehat{\mu}^\top
\left(
\frac{1}{n}\sum_{i = 1}^n \frac{\widehat{g}_i \widehat{g}_i^\top}{1 + \widehat{\lambda}^\top \widehat{g}_i}
\right)
\widehat{\mu}.
\end{equation*}
It follows that \({\vert\widehat{\lambda}\vert\widehat{\mu}^\top
S_n(\widehat{\theta})\widehat{\mu}} 
\leq 
{\widehat{\mu}^\top G_n(\widehat{\theta}) 
(1 + \vert\widehat{\lambda}\vert Z_n)}\), and we have
\begin{equation*}
\vert\widehat{\lambda}\vert 
\left(
\widehat{\mu}^\top
S_n(\widehat{\theta})\widehat{\mu} - Z_n \widehat{\mu}^\top G_n(\widehat{\theta}) 
\right) 
\leq 
\widehat{\mu}^\top G_n(\widehat{\theta}
).
\end{equation*}
From \cref{condition:uniform consistency}, a Taylor expansion of \({G_n(\widehat{\theta})}\) around \({\theta_0}\) yields
\begin{equation*}
G_n(\widehat{\theta}) = 
G_n(\theta_0) + \partial_{\theta}G_n(\theta_0)(\widehat{\theta} - \theta_0) + o_{P}(a_n^{-1}),
\end{equation*}
and we see that \({\widehat{\mu}^\top G_n(\widehat{\theta}
)} = {O_{P}(a_n^{-1})}\) from \cref{condition:asymptotic normality}.
Similarly, from \cref{condition:maximum bound} we have
\begin{equation*}
Z_n \leq \max_{1 \leq i \leq n} \left\vert g\left(X_i, \theta_0\right)\right\vert +
\left(\max_{1 \leq i \leq n} \left\Vert\partial_{\theta}{g}\left(X_i, \theta_0\right)\right\Vert\right) \vert\widehat{\theta} - \theta_0\vert + 
o_{P}(\vert\widehat{\theta} - \theta_0\vert) = o_{P}(a_n).
\end{equation*}
Since \({\widehat{\theta}} \to {\theta_0}\) in probability, there exists a sequence \({\epsilon_n} \to {0}\) such that \({P(\vert\widehat{\theta} - \theta_0\vert} > {\epsilon_n)} \to {0}\), and \({\sup_{\vert\theta - \theta_0\vert \leq \epsilon_n} \Vert S_n(\theta) - S_n(\theta_0)\Vert} \to {0}\) in probability from \cref{condition:consistent covariance estimation}. 
Then for any \({\epsilon} > {0}\),
\begin{equation*}
P\left(\Vert S_n(\widehat{\theta}) - S_n(\theta_0)\Vert > \epsilon\right) 
\leq  
P\left(\sup_{\left\vert\theta - \theta_0\right\vert \leq \epsilon_n} \Vert S_n(\theta) - S_n(\theta_0)\Vert > \epsilon\right) 
+ 
P\left(\vert\widehat{\theta} - \theta_0\vert > \epsilon_n\right),
\end{equation*}
and it follows that \({S_n(\widehat{\theta}) - S_n(\theta_0)} \to {0}\) and \({S_n(\widehat{\theta})} \to {V}\) in probability. 
We write \({\sigma_{\textnormal{min}} + o_{P}(1)} \leq {\widehat{\mu}^\top S_n(\widehat{\theta})\widehat{\mu}} \leq {\sigma_{\textnormal{max}} + o_{P}(1)}\),
with \({0} < {\sigma_{\textnormal{min}}} \leq {\sigma_{\textnormal{max}}}\) denoting the smallest and largest eigenvalues of \({v}\). 
This shows that \({\widehat{\lambda} = O_{P}(a_n^{-1})}\).
Iterating the substitution after \eqref{eq:lambda solution} gives 
\({1 / (1 + \widehat{\lambda}^\top \widehat{g}_i)} = {1 - \widehat{\lambda}^\top \widehat{g}_i + (\widehat{\lambda}^\top \widehat{g}_i)^2 / (1 + \widehat{\lambda}^\top \widehat{g}_i)}\), leading to \({0} = {G_n(\widehat{\theta})} - {S_n(\widehat{\theta})\widehat{\lambda}} + {r_n(\widehat{\theta})}\), where
\begin{equation*}
\vert r_n(\widehat{\theta})\vert
= 
\left\vert
\frac{1}{n}\sum_{i = 1}^n \frac{(\widehat{\lambda}^\top \widehat{g}_i)^2 \widehat{g}_i}{1 + \widehat{\lambda}^\top \widehat{g}_i}
\right\vert \leq 
\max_{1 \leq i \leq n}\vert 1 + \widehat{\lambda}^\top \widehat{g}_i\vert^{-1}
Z_n
\vert \widehat{\lambda}\vert^2
\Vert S_n(\widehat{\theta})\Vert.
\end{equation*}
With \({\max_{1 \leq i \leq n}\vert 1 + \widehat{\lambda}^\top \widehat{g}_i\vert^{-1}} = {O_{P}(1)}\), it follows that \({r_n(\widehat{\theta})} = {o_{P}(a_n^{-1})}\) and 
\begin{equation}\label{eq:lambda}
\widehat{\lambda} = S_n(\widehat{\theta})^{-1} G_n(\widehat{\theta}) + o_{P}(a_n^{-1}),
\end{equation}
where \({S_n(\widehat{\theta})}\) is invertible with probability tending to 1.

Define the empirical likelihood statistic \({l_n(\widehat{\theta})} = {2 a_n^2 n^{-1}\sum_{i = 1}^n \log(1 + \widehat{\lambda}^\top \widehat{g}_i)}\) and apply a Taylor expansion of \({\log(1 + x)}\) to write
\begin{align*}
l_n(\widehat{\theta}) 
&= 
\frac{2a_n^2}{n}
\left(
\widehat{\lambda}^\top \sum_{i = 1}^n \widehat{g}_i - \frac{1}{2} \widehat{\lambda}^\top \sum_{i = 1}^n \widehat{g}_i \widehat{g}_i^\top \widehat{\lambda} + \frac{1}{3}\sum_{i = 1}^n \frac{(\widehat{\lambda}^\top \widehat{g}_i)^3}{(1 + \eta_i)^3}
\right) \\
&=
2 a_n^2 \widehat{\lambda}^\top G_n(\widehat{\theta}) - a_n^2 \widehat{\lambda}^\top S_n(\widehat{\theta}) \widehat{\lambda} + \frac{2}{3} R_n(\widehat{\theta}),
\end{align*}
where \({\vert\eta_i\vert} < {\vert\widehat{\lambda}^\top \widehat{g}_i\vert}\) for all \({i}\) and
\begin{equation*}
\vert R_n(\widehat{\theta})\vert
= 
\left\vert \frac{a_n^2}{n}\sum_{i = 1}^n \frac{(\widehat{\lambda}^\top \widehat{g}_i)^3}{(1 + \eta_i)^3}
\right\vert \
\leq 
a_n^2 \vert\widehat{\lambda}\vert Z_n (1 + \vert\widehat{\lambda}\vert Z_n)^{-3} \vert\widehat{\lambda}^\top S_n(\widehat{\theta}) \widehat{\lambda}\vert
= 
o_{P}(1).
\end{equation*}
From \eqref{eq:lambda}, we have \({l_n(\widehat{\theta})} = {a_n^2 G_n(\widehat{\theta})^\top V^{-1} G_n(\widehat{\theta})} + {o_{P}(1)}\), and it can also be shown that
\begin{equation*}
a_n^2 G_n(\widehat{\theta})^\top V^{-1} G_n(\widehat{\theta}) = a_n^2 (H_n + \widehat{\theta} - \theta_0)^\top M^{-1} (H_n + \widehat{\theta} - \theta_0) + o_{P}(1),
\end{equation*}
where \({H_n} = {WG_n(\theta_0)}\). 
Thus, we can write 
\begin{equation}\label{eq:quadratic approximation}
l_n(\widehat{\theta}) = Q_n(\widehat{\theta}) + o_{P}(1),
\end{equation}
where \({Q_n(\widehat{\theta})} = {a_n^2 (H_n + \widehat{\theta} - \theta_0)^\top M^{-1} (H_n + \widehat{\theta} - \theta_0)}\) is a quadratic approximation to \({l_n(\widehat{\theta})}\).

We now introduce a generic \({q}\)-dimensional function \({h(\theta)}\) with Jacobian matrix \({J}\) from \cref{condition:hypothesis}.
Since \eqref{eq:quadratic approximation} holds for any \({\widehat{\theta}}\) such that \({\widehat{\theta} - \theta_0} = {O_{P}(a_n^{-1})}\), it follows that
\begin{equation}\label{eq:minimization}
\inf_{\theta \in H \cap \overline{K}_n} l_n(\theta) = \min_{\theta \in H \cap \overline{K}_n} Q_n(\theta) + o_{P}(1),
\end{equation}
where \({H} = {\{\theta \in \Theta: h(\theta) = 0\}}\) is the constraint set and \({\overline{K}_n} = {\{\theta: \vert\theta - \theta_0\vert \leq K / a_n\}}\) denotes any sequence of closed balls around \({\theta_0}\) for \({K} > {0}\); see \citet[p.~470]{adimari2010note}.
Thus, we can consider minimizing \({Q_n(\theta)}\), instead of \({l_n(\theta)}\), under the constraint by constructing
\begin{equation*}
L = Q_n\left(\theta\right) + 2a_n^2 h\left(\theta\right)^\top \nu,
\end{equation*}
where \({\nu}\) is a \({q}\)-dimensional Lagrange multiplier. 
Differentiating \({L}\) with respect to \({\theta}\) and \({\nu}\), we obtain
\begin{equation}\label{eq:lagrangian}
\theta - \theta_0 + H_n + MJ^\top \nu = 0 \textnormal{ and } h(\theta) = h(\theta_0) + J(\theta - \theta_0) + o(\vert\theta - \theta_0\vert) = 0.
\end{equation}
Since we only consider solutions \({\tilde{\theta}}\) such that \({\tilde{\theta} - \theta_0} = {O_{P}(a_n^{-1})}\) from \({H_n}\) and \cref{condition:asymptotic normality}, it follows from \eqref{eq:lagrangian} that \({\nu} = {P^{-1}JH_n} + {o_{P}(a_n^{-1})}\), where \({P} = {JMJ^\top}\).
Consequently, we have \({H_n + \tilde{\theta} - \theta_0}  = {MJ^\top P^{-1} J H_n + o_{P}(a_n^{-1})}\), and
\begin{equation}\label{eq:minimization2}
Q_n(\tilde{\theta}) = a_n^2 G_n(\theta_0)^\top A G_n(\theta_0) + o_{P}(1),
\end{equation}
where \({A} = (J W)^\top P^{-1}(J W)\).
It follows from \eqref{eq:minimization} and \eqref{eq:minimization2} that 
\begin{equation*}
\inf_{\theta \in H \cap \overline{K}_n} l_n(\theta) 
= a_n^2 G_n(\theta_0)^\top A G_n(\theta_0) +  o_{P}(1)
\end{equation*}
for some \({K} > {0}\) and \({\overline{K}_n}\).

Now consider hypotheses \({H_j}\) for \({j} = {1, \dots, m}\).
For each \({j}\), there exist \({K_j} > {0}\) and \({\overline{K}_n^j} = {\{\theta: \vert\theta - \theta_0\vert \leq K_j / a_n\}}\) such that \({\inf_{\theta \in H_j \cap \overline{K}_n^j} l_n(\theta)} = {a_n^2 G_n(\theta_0)^\top A_j G_n(\theta_0)} + {o_{P}(1)}\). 
We may take \({K} = {\max\{K_1, \dots, K_m\}}\) and define \({T_{nj}} = {\inf_{\theta \in H_j \cap \overline{K}_n} l_n(\theta)}\).
Then,
\begin{equation*}
T_{nj} = U_{nj}^\top \Sigma_j^{-1} U_{nj} + o_{P}\left(1\right),
\end{equation*}
where \({U_{nj}} = {a_n J_j W G_n(\theta_0)} \to {U_j} \sim {N(0, J_j M J_j^\top)}\) in distribution with \({\Sigma_j} = {J_j M J_j^\top}\). 
Applying the Cr\'amer--Wold device, under \({H_0}\) we have
\begin{equation*} 
T_n = (T_{n1}, \dots, T_{nm}) \to
T \equiv
(U_1^\top \Sigma_1^{-1} U_1, \dots, U_m^\top \Sigma_m^{-1} U_m)
\end{equation*}
in distribution. 
Finally, \({(\Sigma_1^{-1/2} U_1, \dots, \Sigma_m^{-1/2} U_m)}\) has a multivariate normal distribution with mean \({0}\) and correlation matrix \({R}\), where \({R}\) is a \({\sum_{j = 1}^m q_j \times \sum_{j = 1}^m q_j}\) block matrix whose \({k}\)th diagonal matrix is \({I_{q_k}}\) and \({(k, l)}\) off-diagonal matrix is \({\Sigma_k^{-1/2} J_k M J_l^\top \Sigma_l^{-1/2}}\). 
Then, \({T}\) follows a multivariate chi-square distribution with parameters \({m}\), \({q} = {(q_1, \dots, q_m)}\), and \({R}\), i.e.~\({T} \sim {\chi^2(m, q, R)}\).
\end{proof}

\begin{proof}[Proof of \cref{thm:nb}]
We present the proof of \cref{thm:nb} first, followed by the proof of \cref{thm:amc} since it readily follows from \cref{thm:nb}.
We set up some notation and terminology. 
Recall that the bounded Lipschitz metric \({d_\textnormal{BL}}\) between two probability measures \({p}\) and \({Q}\) on \({\mathbb{R}^m}\) for \({m} \in {\mathbb{N}}\) metrizes weak convergence and is defined by
\begin{equation*}
d_{\textnormal{BL}}\left(P, Q\right) = \sup_{f \in \textnormal{BL}_1} \left \vert 
\int f d P - \int f d Q
\right \vert,
\end{equation*}
where \({\textnormal{BL}_1}\) denotes the set of functions \({f: \mathbb{R}^m \mapsto [-1, 1]}\) such that \({\vert f(x) - f(y) \vert \leq \vert x - y \vert}\) for all \({x, y} \in {\mathbb{R}^m}\). 
As a mapping from the common probability space \({(\Omega, \mathcal{F}, P)}\) into the set of probability measures on \({\mathbb{R}^m}\), let \({\widehat{P}_n}\)
be a sequence of random probability measures such that \({\int f d \widehat{P}_n}\) is measurable for any bounded and Lipschitz continuous function \({f}\). 
Then we say that \({\widehat{P}_n}\) converges weakly to \({p}\) in probability if 
\begin{equation}\label{eq:weak_convergence}
\int f d \widehat{P}_n \to \int f d P
\end{equation}
in probability for all \({f} \in {\textnormal{BL}_1}\). 
Also, \eqref{eq:weak_convergence} holds if and only if \({d_{\textnormal{BL}}(\widehat{P}_n, P) \to 0}\) in probability. 
Moreover, if the distribution function of \({p}\) is continuous, it is equivalent to \({d_K(\widehat{P}_n, P) \to 0}\) in probability \citep[see, e.g.][Lemma 2.5]{bucher2019note}, where 
\begin{equation*}
d_K(P, Q) = \sup_{x \in \mathbb{R}^m} \left\vert
P(\{(-\infty, x)\}) - Q(\{(-\infty, x)\})
\right\vert
\end{equation*}
denotes the Kolmogorov distance between \({P}\) and \({Q}\). 

Let \({X_i^*}\) be an independent observation from \({\mathcal{X}_n}\) for \({i} = {1, \dots, n}\).
It can be shown that, for each \({j}\), \({T^*_{nj}} = {n G^*_n(\overline{X})^\top A_j^* G_n^*(\overline{X})} + {o_{P}(1)}\),
where \({G_n^*(\overline{X})} = {n^{-1}\sum_{i = 1}^n g(X_i^*, \overline{X})}\) and
\begin{equation*}
A_j^* 
= 
(J_j {W}^{-1})^\top 
\left(J_j (W^\top {S_n^*}^{-1} W)^{-1} J_j^\top \right)^{-1} 
(J_j {W}^{-1}).
\end{equation*}
We first establish a bootstrap central limit theorem for \({\sqrt{n}G_n^*(\overline{X})}\) as in \citet{singh1981asymptotic}.
Observe that 
\begin{equation*}
E^*\left\{g(X_i^*, \overline{X})\right\} = \frac{1}{n}\sum_{i = 1}^n g(X_i, \overline{X}) = 0,
\end{equation*}
and 
\begin{equation*}
\frac{1}{n}\sum_{i = 1}^n \textnormal{Var}^*\left\{ 
g(X_i^*, \overline{X})
\right\} 
= \frac{1}{n}\sum_{i = 1}^n  
g(X_i, \overline{X}) g(X_i, \overline{X})^\top
= S_n(\overline{X})
= S_n(\theta_0) + o(1)
\end{equation*}
almost surely.
Then \({S_n(\overline{X})} \to {V}\) almost surely by the law of large numbers. 
Applying the Lindeberg--Feller central limit theorem, for any \({\epsilon} > {0}\) we have
\begin{align*}
\frac{1}{n}&\sum_{i = 1}^n E^*\left\{
\vert g(X_i^*, \overline{X})\vert^2 
1\left(\vert g(X_i^*, \overline{X})\vert \geq \epsilon \sqrt{n}
\right)
\right\} \\
&\leq \frac{1}{n}\sum_{i = 1}^n \left\{
\vert X_i - \overline{X}\vert^2 
1\left(\vert X_i - \overline{X}\vert \geq \epsilon \sqrt{n}
\right)
\right\} \to 0
\end{align*}
almost surely.
It follows that \({\sqrt{n}G_n^*(\overline{X})}\) converges weakly to a \({N(0, V)}\) distribution almost surely, i.e.
\begin{equation}\label{eq:bootstrap_clt}
d_K\left(\mathcal{L}\left(\sqrt{n}G_n^*(\overline{X}) \relmiddle | \mathcal{X}_n \right),\ N(0, V)  \right) \to 0
\end{equation}
almost surely.
Next, and let \({s^*_{jk}}\) denote the \({(j, k)}\) component of \({S^*_n}\). 
Then \({E^*(S_n^*)} = {S_n(\overline{X})}\) and, by the law of iterated expectation,
\begin{equation*}
E\left(\Vert S_n^* - S_n(\overline{X})\Vert^2\right) 
\leq 
\sum_{j = 1}^p\sum_{k = 1}^p \frac{1}{n^2}\sum_{i = 1}^n E\left\{(X_{ij} - \overline{X}_j)^2(X_{ik} - \overline{X}_k)^2 \right\} \to 0.
\end{equation*}
This implies that \({S_n^*} \to {V}\) in probability.

It follows from the continuous mapping theorem and \eqref{eq:bootstrap_clt} that \({d_K(\mathcal{L}^*(T_{nj}^*), \mathcal{L}(T_j))} \to {0}\) in probability for each \({j}\). 
Then for every fixed \({\lambda} = {(\lambda_1, \dots, \lambda_m)} \in {\mathbb{R}^m}\), an application of the continuous mapping theorem implies that
\begin{equation}\label{eq:bootstrap_Cramer_wold}
d_K\left(\mathcal{L}^*\left(\lambda^\top T_n^* \relmiddle | \mathcal{X}_n\right),\ \mathcal{L}(\lambda^\top T) \right) \to 0
\end{equation}
in probability. 
From the subsequential property of convergence in probability \citep[Theorem 9.2.1]{dudley2002}, there exists a subsequence such that \eqref{eq:bootstrap_Cramer_wold} holds almost surely along the subsequence. 
Then the Cr\'amer--Wold device implies that \({T_n^*}\) converges weakly to \({T}\) almost surely along the subsequence. 
Another application of the subsequential argument shows that 
\begin{equation*}
d_K\left(\mathcal{L}\left(T_n^* \relmiddle | \mathcal{X}_n\right),\ \mathcal{L}(T) \right) \to 0
\end{equation*}
in probability. Finally, \({T_n} \to {T}\) in distribution under \({H_0}\) and the result follows from the continuity of \({T}\).
\end{proof}

\begin{proof}[Proof of \cref{thm:amc}]
Since \({S_n(\widehat{\theta})} \to {V}\) in probability, we have \({\widehat{A}_j} \to {A_j}\) in probability for \({j} = {1, \dots, m}\) and the continuity of the characteristic function of the normal distribution implies that
\begin{equation*}
d_K\left(
\mathcal{L}\left( U_n \relmiddle | \mathcal{X}_n \right),
N(0, V) \right)
\to
0
\end{equation*}
in probability. 
For any \({\lambda} = {(\lambda_1, \dots, \lambda_m)} \in {\mathbb{R}^m}\), it follows from the continuous mapping theorem that
\begin{equation*}
d_K\left(
\mathcal{L}\left(U_n^\top
\left(\sum_{j = 1}^m \lambda_j \widehat{A}_j \right)
U_n \relmiddle | \mathcal{X}_n \right),\ 
\mathcal{L}
\left(
U^\top
\left(\sum_{j = 1}^m \lambda_j A_j \right)
U\right)
\right)
\to
0
\end{equation*}
in probability. Then the Cr\'amer--Wold device and the subsequential argument applied in \eqref{eq:bootstrap_Cramer_wold} complete the proof.
\end{proof}

\end{document}